\documentclass[11pt,oneside,reqno]{amsart}
\usepackage{url}
\usepackage{amsmath}
\usepackage{graphicx}
\usepackage{amssymb}
\usepackage{bbm}
\usepackage{bm}
\usepackage{tikz}
\usepackage{enumerate}
\usepackage{cite}
\usepackage{natbib}
\setcitestyle{authoryear, round}
\usepackage{hyperref}
\usepackage{graphicx}
\usepackage{algorithm}
\usepackage{algorithmicx}
\usepackage[title]{appendix}
\usepackage{algpseudocode}
\usepackage{float}
\usepackage{numprint}
\usepackage{amsaddr}
\usepackage[margin=2.5cm]{geometry}

\newcommand{\likelil}{\mathcal{L}}
\newcommand{\Var}{\textrm{Var}}

\hypersetup{
           breaklinks=true,   
           colorlinks=false,   
           pdfusetitle=true,  
        }

\begin{document}
\bibliographystyle{abbrvnat}
\setcitestyle{authoryear, round}

\title[Article Title]{Novel Bayesian algorithms for ARFIMA long-memory processes: a comparison between MCMC and ABC approaches}

\author{{James Cohen Gabor}, {Clara Grazian}}
\address{University of Sydney, Australia; \\ARC Training Centre in Data Analytics for Resources and Environments}
\email{{james.gabor@sydney.edu.au, clara.grazian@sydney.edu.au}}

\begin{abstract}This paper presents a comparative study of two Bayesian approaches—Markov Chain Monte Carlo (MCMC) and Approximate Bayesian Computation (ABC)—for estimating the parameters of autoregressive fractionally-integrated moving average (ARFIMA) models, which are widely used to capture long-memory in time series data. We propose a novel MCMC algorithm that filters the time series into distinct long-memory and ARMA components, and benchmarked it against standard approaches. Additionally, a new ABC method is proposed, using three different summary statistics used for posterior estimation. The methods are implemented and evaluated through an extensive simulation study, as well as applied to a real-world financial dataset, specifically the quarterly U.S. Gross National Product (GNP) series. The results demonstrate the effectiveness of the Bayesian methods in estimating long-memory and short-memory parameters, with the filtered MCMC showing superior performance in various metrics. This study enhances our understanding of Bayesian techniques in ARFIMA modeling, providing insights into their advantages and limitations when applied to complex time series data.
\end{abstract}


\maketitle
\section{Introduction}\label{intro}

In many fields, from natural science to economics, the complex interactions characterizing observable processes have led to the development of increasingly sophisticated models. Many of these advancements involve models capable of capturing intricate correlation structures. One such class of models addresses long-memory, a phenomenon where the autocorrelations of a time series decay at a hyperbolic rate. A widely used model for capturing long-memory is the autoregressive fractionally-integrated moving average (ARFIMA) model, an extension of the ubiquitous autoregressive moving average (ARMA) model, whose correlation structure is characterized by geometric decay. The appeal of a model with a stochastic structure that allows for non-negligible autocorrelations at large lags has led to widespread research and application of ARFIMA models \citep{granger1980introduction,koop1997bayesian,hauser1999maximum,robinson2003time,doornik2004inference}.

In research on modeling long-memory processes using ARFIMA models, most methods employ standard parametric or semi-parametric approaches \citep{RN41,geweke1983estimation,sowell1992maximum,beran1995maximum,baillie1996fractionally,doornik2004inference,shimotsu2005exact,chan2006estimation}, with only a few studies focusing on Bayesian methods. Among the proposed Bayesian approaches, \cite{RN3} is notable for introducing a Markov Chain Monte Carlo (MCMC) methodology that utilizes a Metropolis-within-Gibbs sampler. This methodology, like others that directly evaluate the ARFIMA likelihood, requires extensive repetitions of matrix operations and inversions, which is computationally demanding. However, it does not require approximating the ARFIMA process with a more easily tractable likelihood. On the other hand, \cite{RN59} applied an importance sampler to an ARMA model, with the order chosen to approximate the long-memory of an ARFIMA process. This method is faster compared to MCMC but depends on the accuracy of the ARMA process in approximating a long-memory ARFIMA process. More recently, \cite{RN56} utilized parallel GPU computing to implement a sequentially adaptive Bayesian learning algorithm on ARFIMA models. This approach can be parallelized, unlike other MCMC methods where the chain directly depends on the previous value. However, the method relies on the availability of scalable GPU systems and still requires a significant amount of computational time. More recently, \cite{grazian2024bayesian} proposed a Bayesian semi-parametric approach to estimate the parameter of long-memory for general long-memory processes, reducing the parametric assumptions. 

Currently, there is no generally accepted ``best'' Bayesian approach for analyzing ARFIMA processes, and a direct comparison among available methodologies is still lacking in the literature. One of the challenges in estimating ARFIMA processes using a Bayesian framework is the increased computational complexity associated with evaluating the likelihood function and simulating from the posterior distributions. This computational burden is exacerbated by the fact that ARFIMA models are often theoretically justified only for datasets with a large number of observations. For smaller datasets, even when long-memory is suspected, analysis is often better performed using ARMA models rather than ARFIMA models \citep{RN58}. This effectively necessitates having a sufficiently large series to justify fitting an ARFIMA model, meaning any Bayesian ARFIMA analysis will involve handling large datasets. Nevertheless, we consider the Bayesian approach to be beneficial for long-memory analysis due to its handling of uncertainty around target parameters. Additionally, the presence and strength of long-memory may be more effectively determined through the analysis of a distribution rather than a point estimate - such an analysis is naturally suited to the Bayesian paradigm.

Several studies have highlighted the possibility of model misspecification when using ARFIMA processes, especially when the long-memory component may be ``spurious'' and mistaken for a higher-order ARMA process, a series exhibiting regime switching, or a non-stationary ARIMA process \citep{RN69,RN61,RN68}. Similar issues have been observed in predictive analyses using ARFIMA models \citep{RN67}. Given the estimation challenges many models face regarding long-memory, it is worthwhile to explore and compare the effectiveness of alternative Bayesian methodologies, whose density estimates may better determine ARFIMA parameters.

The goal of this paper is to provide a comparative study of MCMC and approximate Bayesian computation (ABC) methods for estimating ARFIMA parameters, using a large simulation study, and a real data application from finance. One MCMC algorithm, derived from the literature, serves as a benchmark, while we propose a novel MCMC algorithm and a novel ABC algorithm, implemented in three different versions, comparing their performances. A Bayesian forecasting method is also applied to the sampling methods, using the full ARFIMA posterior predictive distribution. By comparing these distinct approaches, we aim to enhance our understanding of the benefits and drawbacks of Bayesian methods for ARFIMA analysis.

The benchmark MCMC method is based on the Metropolis-within-Gibbs algorithm from \cite{RN3}, while our new proposal is based on filtering the target series into a pure long-memory component and an ARMA process component. This allows for the separate estimation of long-memory and short-memory parameters with an additional Gibbs sampler step. The proposed MCMC filtering is inspired by the result that an ARFIMA$(p,d,q)$ model can be differenced or ``whitened'' into distinct ARMA$(p,q)$ and ARFIMA$(0,d,0)$ series \citep{RN85}.

Additionally, we compare the exact MCMC algorithms with a new approximate method based on an ABC rejection sampling algorithm. For the ABC approach, three versions are considered, each varying the summary statistic used to accept posterior samples: the first two methods use the periodogram as an estimator of the spectral density, and the second employs the log-periodogram regression model proposed by \cite{RN32}.

The remainder of the paper is organized as follows: Section \ref{background} introduces ARFIMA models, long-memory processes, and the theory behind ABC and MCMC frameworks. Section \ref{section3} presents the Bayesian framework for ARFIMA models, detailing the proposed MCMC and ABC algorithms. Section \ref{section4} reports on a simulation study comparing the proposed MCMC and ABC algorithms. Section \ref{section5} applies the developed methods to an economic dataset of quarterly U.S. Gross National Product (GNP) records. Finally, Section \ref{section6} concludes the paper with a discussion.

\section{Background}\label{background}

\subsection{ARFIMA and long-memory}

A time series $\bm{y}_n=(y_1,\dots,y_n)$ is a realisation of an \\ARFIMA$(p,d,q)$ process if it can be expressed as
\begin{align}
    \phi(B)y_t  (1-B)^{d} &=  \theta(B)w_t, \label{q2:arima}
\end{align}
where $B$ is the backshift operator $B^t y_i = y_{i-t}$, for $i \geq t$. The polynomials $\phi(x)= \phi_1 x - \phi_2 x^2 - \dots - \phi_p x^p$ and $\theta(x)=1+\theta_1 x + \theta_2 x^2 + \dots + \theta_q x^q$ describe the autoregressive and moving-average components, respectively. Here, $\phi=(\phi_1,\dots,\phi_p)$ and $\theta=(\theta_1,\dots,\theta_q)$ are the ARMA parameters, with $p$ and $q$ being positive integers representing the ARMA orders. The parameter $d$ represents the order of fractional differencing, and $w_t$ are independent and identically distributed (i.i.d.) random variables. We assume that $w_t \sim N(0,\sigma^2)$. In this paper, for ease of notation, we consider only the case of ARMA orders up to $p=1$ and $q=1$; therefore we refer to the ARMA parameters as $\phi=\phi_1$ and $\theta=\theta_1$. Extensions to higher orders are straightforward. 

The process is stationary and invertible when $d \in (-0.5,0.5)$ and the roots of the ARMA polynomials $\phi(x)$ and $\theta(x)$ lie outside the unit circle. For an ARFIMA$(1,d,1)$ process, this implies $\phi \in (-1,1)$ and $\theta \in (-1,1)$. Note that the range of $d$ for a stationary-invertible ARFIMA process can be extended to $d \in (-1,0.5)$, instead of the range originally proposed in \cite{RN41}; see, for example, \cite{RN53}. In the special case where $d=0$, the process reverts to a standard ARMA process. 

For a stationary process, the ARFIMA process can also include a mean term $\mu$ 
\begin{align}
    \phi(B)(y_t - \mu)  (1-B)^{d} &=  \theta(B)w_t.\notag
\end{align}
Often, the series is ``demeaned'' by subtracting the sample mean from each point \citep{RN54}; this practice is applied here, so all later references to an ARFIMA model assume the form in \eqref{q2:arima}. 

A weakly stationary (second-order stationary) stochastic process with spectral density $f(\lambda)$, and frequency $\lambda \in (-\pi,\pi)$, exhibits long-memory if 
\begin{align}
    \lim_{\lambda \rightarrow 0}f(\lambda) \sim \ell_1(\lambda) |\lambda|^{-2d},\ d \in (0,0.5), \label{longmem_spectral_def}
\end{align}
where $\ell_1(\cdot)$ is a slowly varying function. Specifically, for some  $\ell_1 : (a,\infty) \rightarrow \mathbb{R}^+$ and $a \geq 0, b\in \mathbb{R}^+$, we have $\ell_1(bx) \sim \ell(x),\ x\rightarrow \infty$. This contrasts with the case of short-memory, which is present in stationary ARMA models with $d=0$, where
\begin{align}
    \lim_{\lambda \rightarrow 0}f(\lambda) \sim \ell_1(\lambda) \rightarrow c ,\ c\in \mathbb{R}^+.\notag
\end{align}
In other words, short-memory in ARMA models occurs when $d=0$ and the spectral density approaches a constant as $\lambda \rightarrow 0$. The case where $d=0$ but the spectral density does not converge to a constant function is described as an intermediate-memory process \citep{RN66}. The parameter $d$ in \eqref{longmem_spectral_def} is commonly referred to as the long-memory parameter.

An alternative definition for long-memory can be presented in the time domain. If a weakly stationary stochastic process has an autocovariance function (ACVF) $\gamma(h)$ at lag $h\in \mathbb{Z}$, and $\ell_2$ is a slowly varying function, then the process exhibits long-memory if
\begin{align}
    \lim_{h\rightarrow \infty}\gamma(h) \sim \ell_2(h) h^{2d-1},\ \ \ d\in (0,0.5)\notag.
\end{align}
A result from this definition is that the sum of autocorrelations is unbounded for long-memory processes but bounded for short-memory processes:
\begin{align}
\sum_{h=-\infty}^{\infty}|\gamma(h)| &\rightarrow \infty,\ \ \ d\in(0,0.5) \text{ (long-memory)},\notag\\
\sum_{h=-\infty}^{\infty}|\gamma(h)| &\rightarrow c,\ \ \ d=0 \text{ (short memory)}\notag.
\end{align}

The case where $d\in (-0.5,0)$ is commonly described as an antipersistent process. For antipersistent processes, the spectrum does not exhibit long-memory - in fact, the corresponding sum of autocorrelations converges. The case where $d \in (-1,-0.5)$ is more rare in practice, and it usually arises from `overdifferencing' a series. It can indicate a model misspecification \citep{RN53,beran2017statistics,hassler2018time}. While the primary use of ARFIMA models is to capture long-memory phenomena, which occur only for $d\in (0,0.5)$, the full stationary range of $d\in(-0.5,0.5)$ is often considered. Therefore, we will assume that the domain of $d$ is $(-0.5,0.5)$ in this work as well, and we will refer to $d$ as ``long-memory parameter'', even if it can indicate short-memory when $d=0$ and antipersistence when $d \in (-0.5,0)$. 

\subsection{Algorithms for Bayesian analysis} \label{sub:algo_for_bayes}

In a Bayesian framework, model inference is carried out using the posterior distribution, which describes the model parameter $\psi \in \Psi$, conditional on the observed data $\bm{y}_n$; for ARFIMA models the parameter is defined as $\psi = (d,\phi,\theta,\sigma^2)$. The form of the posterior distribution requires the specification of the likelihood function, $\likelil(\psi;\bm{y}_n)$. The likelihood function derived from model \eqref{q2:arima} under the assumption of zero-mean Gaussian realisations has the form
\begin{small}
\begin{align*}
    \likelil&(\psi;\bm{y}_n) = (2\pi \sigma^2)^{-n/2} |\Sigma_n|^{-1/2} \cdot \exp\bigg\{-\frac{1}{2\sigma^2} \bm{y}_n^T \Sigma_n^{-1} \bm{y}_n\bigg\},
\end{align*}
\end{small}
where $\sigma^2 \Sigma_n$ is the covariance matrix and elements $\{\Sigma_{ij}\},\ i,j=1,\dots,n$ are given by ARFIMA ACVFs $\gamma(|i-j|)$. The likelihood function is combined with the prior beliefs about the model parameters through a prior distribution $\pi(\psi)$ to obtain the posterior distribution:
\begin{align}
\pi(\psi | \bm{y}_n) &= \frac{\likelil(\psi;\bm{y}_n) \pi(\psi)}{\int_\Psi \likelil(\psi;\bm{y}_n)\pi(\psi)d\psi}.\notag
\end{align}

Given $(d, \phi, \theta)$ are parameters defined in compact spaces, default choices for their prior distributions are $d\sim Unif(-0.5,0.5)$, $\phi \sim Unif(-1,1)$, and $\theta \sim Unif(-1,1)$, and it is reasonable to use a conjugate inverse-gamma prior for $\sigma^2 \sim \text{IG}(\alpha, \beta)$. The posterior distribution for an ARFIMA$(1,d,1)$ model can then be written as
\begin{align}
    \pi(d,\phi,\theta,\sigma^2 &| \bm{y}_n) \propto (2\pi \sigma^2)^{\frac{n}{2}-\alpha-1} | \Sigma_n|^{\frac{1}{2}} \nonumber \\
    \quad &\cdot \exp\bigg\{-\frac{1}{2\sigma^2}[ \bm{y}_n^T \Sigma_n^{-1} \bm{y}_n + 2\beta ] \bigg\}. \label{eq:postall}
\end{align}

This posterior distribution is not known in closed form due to the dependence of $\Sigma_n$ on $(d,\phi,\theta)$, and requires numerical methods for approximation. Two common categories of methods for estimating the posterior distribution are Markov Chain Monte Carlo (MCMC) and approximate Bayesian computation (ABC) methods. Generally, MCMC methods rely on the ability to evaluate the posterior distribution $\pi(\psi | \bm{y}_n)$ up to a normalizing constant, and thus require repeated evaluation of the likelihood function $\likelil(\psi;\bm{y}_n)$. These methods are called exact because they aim to sample values from the exact posterior distribution. 

Often, the likelihood is either costly to evaluate. In such cases, ABC methods can be used. Unlike MCMC, ABC methods, often referred to as ``likelihood-free'', bypass repeated likelihood evaluation by simulating data from the specified model and accepting posterior samples based on the similarity between the simulated and observed data. 

In this work, we propose a novel MCMC algorithm to estimate ARFIMA models and then focus on the effectiveness of ABC methods in estimating ARFIMA models compared to MCMC and frequentist approaches. In the following section, we describe the framework of MCMC and ABC methods that will be used for sampling from the posterior distribution of the parameters of an ARFIMA model. 

\subsection{MCMC and ABC algorithms}

Two commonly used MCMC procedures for drawing samples from a target posterior distribution $\pi(\psi | \bm{y}_n)$ are the Metropolis-Hastings (MH) algorithm and the Gibbs sampler (GS). Each algorithm returns a sequence of draws $\psi_1,\dots,\psi_M$ after $M$ iterations. The appropriate value for $M$ is problem-specific and usually requires a `burn-in' period, where a number of initial simulations are discarded before the resulting Markov chain can be considered to have converged to the stationary target distribution (the posterior distribution). We refer the reader to \cite{gilks1995markov} and \cite{robert1999monte} for a comprehensive presentation of a wide range of MCMC methods, including both MH and GS. 

MH requires a proposal distribution $p(\cdot)$, which may depend on the previously accepted value, from which to draw samples. To perform the acceptance step for the proposed value, it is necessary that the target posterior distribution $\pi(\psi | \bm{y}_n)$ can be analytically evaluated up to a normalizing constant.

GS is a special case of the MH algorithm, particularly useful when the conditional distributions of a partitioned set of parameters can be directly sampled from. Suppose the random variable can be partitioned into $v$ components $\psi = (\psi^{(1)},\dots,\psi^{(v)})$, and that the conditional densities (full conditionals) of each partition given the remaining variables can be sampled from. The GS iteratively generates sample draws from each partition $\psi^{(g)},\ g=1,\dots,v$, and takes the resulting vector of values as a sample from the full target posterior $\pi(\psi | \bm{y}_n)$. It is often the case, particularly for non-trivial Bayesian models, that the full conditionals are not known for all components of $\psi$. In these situations, some samples may be drawn from the conditionals using MH, while others with known full conditionals can be drawn using a GS step. This combination of MH and GS is commonly referred to as Metropolis-within-Gibbs.

ABC is a class of methods typically used for estimating parameters in models with computationally intractable likelihood functions—that is, likelihood functions that are not available in closed form or are costly to evaluate. In ABC, the evaluation of the likelihood function is replaced by simulation from it. In this work, we consider the rejection sampling algorithm. The general rejection sampling algorithm requires the specification of a proposal distribution from which to generate parameter values, e.g. the prior distribution $\pi(\psi)$, as well as the ability to sample from the likelihood function $\likelil(\psi;\bm{y}_n)$. 

In ABC, new simulated datasets are generated using proposed parameter values and then compared to the observed dataset, typically using summary statistics. The comparison is made against a tolerance level, which represents the error the experimenter is willing to accept. Two important decisions are the choice of summary statistics $s=S(\bm{y}_n)$ and the choice of the tolerance level $\epsilon$. Summary statistics are used to reduce the dimensionality of the data and capture its essential features, making it feasible to compare simulated data with observed data. The choice of summary statistics can significantly affect the efficiency and accuracy of the inference, as they should effectively summarize the information relevant to the parameters of interest. Additionally, the tolerance level determines the maximum allowable discrepancy between the simulated and observed summary statistics. A smaller tolerance level results in a more accurate approximation of the posterior distribution but requires more simulations, while a larger tolerance level reduces computational cost but may lead to less precise estimates. Balancing these choices is essential to ensure the effectiveness of the ABC method in approximating the posterior distribution.

In ABC rejection sampling, proposed parameters values are accepted if $\| s_\text{obs} - s_\text{sim} \| < \epsilon$, where $s_\text{obs}$ and $s_\text{sim}$ are the summary statistics computed on the observed and simulated datasets respectively, and $\| \cdot \|$ is a distance, for example the Euclidean distance. Often, $\epsilon$ is set as a target quantile of the distance $\| s_\text{obs} - s_\text{sim} \|$ based on $M$ simulations. After having simulated $M$ proposed parameters, and computed their corresponding summary statistics $s_{\text{sim}}$, only the proposed values leading to the $q_{\epsilon}$-smallest quantile are accepted to be part of the sample approximating the posterior distribution \citep{RN63}.

In this setting, with observed data $\bm{y}_n$ and observed summary statistic $s_\text{obs} = S(\bm{y}_n)$, the ABC rejection algorithm samples from the distribution
\begin{align*}
    \tilde{\pi}(\psi | s_\text{obs}) \propto \int_\mathcal{S} \mathbb{I}\{\|s_\text{obs} - s\| < \epsilon\} \likelil(s_\text{obs} ; \psi)\pi(\psi) ds,  
\end{align*}
where $\mathbb{I}$ is the indicator function. As $\epsilon \rightarrow 0$, the density approaches the posterior of $\psi$ given $s_\text{obs}$, $\pi(\psi | s_\text{obs})$.

If the function $S(\cdot)$ is sufficient for $\bm{y}_n$ (ideally minimally sufficient), then $\pi(\psi | s_\text{obs}) = \pi(\psi | \bm{y}_n)$, and the ABC algorithm would be exact for $\epsilon \rightarrow 0$. When a low-dimensional sufficient statistic is unavailable, the summary statistics $S(\cdot)$ should be carefully chosen to minimise the differences between $\pi(\psi | s_\text{obs})$ and $\pi(\psi | \bm{y}_n)$. The trade-off in choosing an insufficient summary statistic lies between retaining an appropriate amount of information about the observed data and avoiding the curse of dimensionality as the dimension of $s_{\text{obs}}$ increases. In situations where the data and summary statistic can be reasonably assumed to follow standard distributions, the loss incurred by using an insufficient summary statistic can be sometimes quantified \citep{RN64}. However, in cases where the model is complex, and the summary statistic has an unknown distribution, determining the exact loss of accuracy is more challenging. Thus the ABC rejection algorithm, in contrast to most MCMC methods, is only approximate in the sense that it samples from an approximate version of the posterior distribution. 

\section{Bayesian Methods for Estimating ARFIMA Models} \label{section3}

Here, we propose two classes of algorithms to estimate ARFIMA processes within a Bayesian framework. First, we consider exact estimation via MCMC simulations, and second, approximate estimation via ABC. We focus on ARFIMA models with orders up to $p,q \leq 1$; specifically, the notation will be based on an ARFIMA$(1,d,1)$ model with parameters $\psi = (d,\phi,\theta,\sigma^2)$; however, the algorithms can be easily extended to higher orders. Estimations for ARFIMA$(1,d,0)$ and ARFIMA$(0,d,1)$ models are obtained as simplified versions of the proposed algorithms by omitting the relevant $\phi$ or $\theta$ parameter. 

Before presenting the algorithms, we investigate some potential difficulties for inferential procedures when estimating the parameters of ARFIMA models. 

\subsection{Difficulties in estimating ARFIMA parameters}
\label{sub:difficulties}

The possibility of misspecification in the estimation of the long-memory parameter $d$ has been well documented, especially in the presence of non-zero ARMA parameters \citep{RN5,RN70}. Particularly impactful is the relationship between the AR parameter $\phi$ and the long-memory parameter $d$, where large values of $\phi$ can be `mistaken' for long-memory. In fact, long-memory can be represented by an ARMA model when 
$p$ and $\phi$ are large enough \citep{RN59}.

\begin{figure}[h]
\begin{center}
\includegraphics[width=10cm,height=7cm]{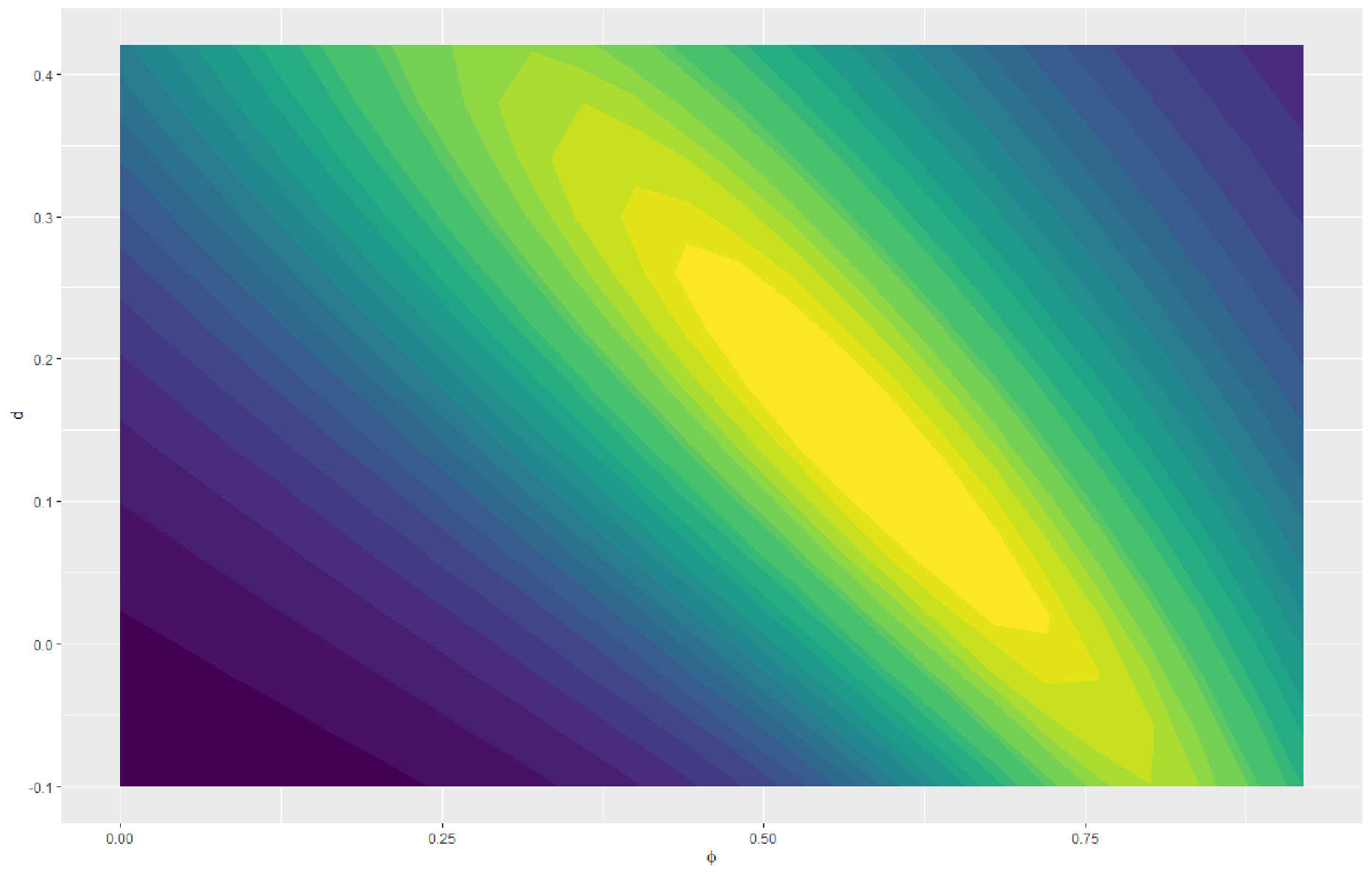}
\caption{Contour plot across a grid of $(d,\phi)$ for the log-posterior density of the parameters of an ARFIMA$(1,d,1)$ process, with true values $d=0.2, \phi=0.5, \theta=0.5$. Larger values of log-density are in yellow, and smaller values are in blue.} \label{plots:arfima1d1_posterior_d0.2_phi0.5_theta0.5}
\end{center}
\end{figure}

Within the context of drawing samples from the posterior density $\pi(d,\phi,\theta,\sigma^2|\bm{y}_n)$, the relationship between $d$ and $\phi$ can be illustrated by examining the joint posterior density of $(d,\phi)$ shown in Figure \ref{plots:arfima1d1_posterior_d0.2_phi0.5_theta0.5}. The density shows that, for an ARFIMA$(1,d,1)$ process, there is a clear negative correlation between $d$ and $\phi$, with high posterior density values concentrated around values where $d$ is smaller than the true value and $\phi$ is larger than the true value.

An issue regarding the estimation of $(d,\phi,\theta)$ is that larger values for $\phi$ can obfuscate the true value of $d$, leading to a downward bias in the marginal posterior density for $d$. The consequence is that, for large $\phi$, the posterior may be centered on a value for $d$ that is lower than the true value from which the data was generated. This effect of $\phi$ being `mistaken' for long-memory is particularly noticeable when $\phi$ is large and $d$ is small (roughly $d < 0.2$) \citep{RN5}. While this is a characteristic of the likelihood function of an ARFIMA model, and it affects all methods based on the likelihood function including Bayesian procedures, we will investigate whether the bias can be mitigated using the proposed approaches. 

\subsection{MCMC Framework for ARFIMA models}
\label{sub:MCMC}

Here, we present the algorithms for generating posterior samples for the parameters of an   ARFIMA$(1,d,1)$ model using MCMC methodologies. Two MCMC algorithms are implemented: the first closely follows the work of \cite{RN3} and will be used as a benchmark; the second introduces a novel approach for estimating the ARMA and long-memory parameters separately by filtering the ARFIMA$(1,d,1)$ process into its respective ARMA$(1,1)$ and ARFIMA$(0,d,0)$ components. In all schemes, we consider the time series $\bm{y}_n$ to be a zero-mean stationary time series; this approach differs slightly from other methodologies that estimate $\mu$. We omit $\mu$ as we are primarily interested in the estimation of the long- and short-memory components. Additionally, since we only consider stationary time series, $\bm{y}_n$ can be easily demeaned into a zero-mean series by subtracting the series mean. 

\subsubsection{ARFIMA MH-within-GS Sampling Scheme - Simultaneous MCMC}
\label{subsub:simultaneousMH}

Evaluation of the posterior distribution $\pi(d,\phi,\theta,\sigma^2 | \bm{y}_n)$  requires calculating the theoretical autocovariances $\{\Sigma_{ij}\}$, $i,j=1,\dots,n$ of an ARFIMA$(1,d,1)$ process. For the case of $p,q\leq 1$, the exact theoretical ACVF can be calculated, as derived by \cite{RN4} (see Chapter 3.2.7):
\begin{align}
    \gamma(h) &= \frac{1}{\phi(\phi^2-1)} \big[\theta C(d,-h,-\phi) + \notag\\
    &\ \ \ \ \theta C(d,2-h,-\phi) + (1+\theta^2) C(d,1-h,-\phi)\big], \label{eq:gammah}
\end{align}
with 
\begin{align}
    C(d,h,\rho) &= \frac{\gamma_0(h)}{\sigma_2} \big[\rho^{2} F(d+h,1,1-d+h,\rho) + \notag\\
    &\ \ \ \ \ F(d-h,1,1-d-h,\rho) - 1\big], \label{eq:Cfun}
\end{align}
where $F(a,b,c,x)$ is the Gaussian hypergeometric function. 

As mentioned, the full conditionals required for a Gibbs sampler cannot be easily derived. This is because the parameters $(d,\phi,\theta)$ are embedded within the correlation matrix $\Sigma_n$ through the ACVF and thus unavailable in closed form. Instead, a MH step can be implemented to sample from $\pi(d,\phi,\theta | \bm{y}_n)$, which has a simpler form.  

The marginal posterior density of $(d,\phi,\theta)$ can be derived by integrating $\sigma^2$ out:
\begin{align}
    \pi(d,\phi,\theta | \bm{y}_n) &\propto \int_{0}^{\infty} \pi(d,\phi,\theta,\sigma^2 | \bm{y}_n) d\sigma^2\notag\\
    &= | \Sigma_n|^{-\frac{1}{2}} \int_{0}^{\infty} (2\pi \sigma^2)^{\frac{n}{2}-\alpha-1}  \cdot \notag\\
    &\ \ \ \ \ \ \exp\bigg\{-\frac{1}{2\sigma^2}[ \bm{y}_n^T \Sigma_n^{-1} \bm{y}_n + 2\beta ] \bigg\} d\sigma^2 \\
    &\propto |\Sigma_n|^{-\frac{1}{2}} \big(\bm{y}_n^T \Sigma_n^{-1} \bm{y}_n + 2\beta \big)^{-\frac{n}{2} - \alpha}.
    \label{eq:integratedpost}
\end{align}
The marginal posterior density can be used as the target distribution of the MH step. 

Once values from $\pi(d,\phi,\theta|\bm{y}_n)$ are obtained, they can then be used to sample values for $\sigma^2$ from its full conditional $\pi(\sigma^2 | \bm{y}_n, d, \phi, \theta)$: 
\begin{align}
    \sigma^2 | \bm{y}_n, d, \phi,\theta \sim \text{IG}\bigg(\frac{n}{2} + \alpha, \frac{1}{2}(\bm{y}_n^T \Sigma_n^{-1} \bm{y}_n + 2\beta)\bigg).
    \label{sig2_posterior} 
\end{align}

Full steps of the procedure are detailed in Algorithm \ref{algo:mcmc_method1}.
An algorithm for the parameters of an ARFIMA$(1,d,0)$ or ARIMFA$(0,d,1)$ model can be easily obtained with a reformulation of the covariance matrix $\Sigma_n$. 

\begin{algorithm}[h]
\caption{ARFIMA$(1,d,1)$ Simultaneous MCMC } \label{algo:mcmc_method1}
Given series $\bm{y}_n$, hyperparameters $\alpha, \beta$, proposal $p(\cdot)$, and initial values $(d_0, \phi_0, \theta_0)$,
\begin{algorithmic}
\For{iteration $i=1,\dots,M$}
    \medskip
    \State \textbf{MH Random Walk Step}:
    \State Draw proposal values $(\tilde{d},\tilde{\phi},\tilde{\theta}) \sim p(\cdot | d_{i-1},\phi_{i-1},\theta_{i-1})$.
    \State Calculate the acceptance probability 
        \begin{align*}
            \alpha = \min\bigg\{0, \frac{ \pi(\tilde{d},\tilde{\phi},\tilde{\theta} | \bm{y}_n)}{\pi(d_{i},\phi_{i},\theta_{i} | \bm{y}_n)} \cdot \frac{p(d_{i},\phi_{i},\theta_{i} | \tilde{d},\tilde{\phi},\tilde{\theta})}{p(\tilde{d},\tilde{\phi},\tilde{\theta} | d_{i},\phi_{i},\theta_{i})}\bigg\}.
        \end{align*}
    \State Simulate $w \sim \text{Bernoulli}(\alpha)$
    \If{$w = 1$}
        \State Set $(d_{i},\phi_{i},\theta_{i}) = (\tilde{d},\tilde{\phi},\tilde{\theta})$.
    \Else
        \State Set $(d_{i},\phi_{i},\theta_{i}) = (d_{i-1},\phi_{i-1},\theta_{i-1})$.
    \EndIf
    \medskip
    \State \textbf{GS Step}:
    \State Calculate the ARFIMA$(1,d,1)$ covariance matrix  $\Sigma_{n,i}$ using $(d_{i},\phi_{i},\theta_{i})$.
    \State Draw $\sigma_i^2 | d_{i}, \phi_{i}, \theta_{i},\bm{y}_n \sim \text{IG}\big(\frac{n}{2} + \alpha,\frac{1}{2}(\bm{y}_n^T\Sigma_{n,i}^{-1} \bm{y}_n + 2\beta) \big)$.
\EndFor

\medskip
\State \textbf{Output}:
\State A set of MCMC samples $\{(d_i,\phi_i, \theta_i,\sigma^2_i) : i=1,\dots,M\}$ from $\pi(d,\phi,\theta,\sigma^2|\bm{y}_n)$.

\end{algorithmic}
\end{algorithm}

\subsubsection{ARFIMA MH-within-GS Sampling Scheme - Filtered MCMC} \label{filtered_mcmc_section}

Inspired by the frequentist work of \cite{RN41} and \cite{RN29}, we propose a novel MCMC algorithm by filtering the ARFIMA series into separate ARFIMA$(0,d,0)$ and ARMA$(1,1)$ processes, enabling sequential sampling and acceptance of $d$ and $(\phi,\theta)$. This filtering approach aims to reduce the influence of the ARMA process on the estimation of the long-memory parameter $d$, and vice-versa, to mitigate the inferential difficulties described in Section \ref{sub:difficulties}. For ARFIMA$(1,d,0)$ or ARFIMA$(0,d,1)$ models, the series can be similarly filtered into an ARFIMA$(0,d,0)$ and ARMA$(1,0)$ or ARMA$(0,1)$ process, respectively.

First, we can notice that $\bm{y}_n$ can be filtered into an ARFIMA$(0,d,0)$ process $\bm{z}_n$ using 
$$
z_t = \frac{\phi(B)}{\theta(B)} y_t, \qquad t=1, \ldots,n.
$$
Similarly, the series $\bm{y}_n$ can be filtered into an ARMA$(1,1)$ series $\bm{u}_n$ using
$$
u_t= (1-B)^{d}y_t, \qquad t=1, \ldots,n.
$$

Once the filtered series are obtained, the MCMC algorithm can be defined by two iterative steps, one targeting $\pi(d | \bm{z}_n, \phi, \theta)$, and one targeting $\pi(\phi,\theta | \bm{u}_n, d)$. If $\bm{y}_n$ follows an ARFIMA$(1,d,1)$ process, then for $\pi(d | \bm{z}_n, \phi, \theta)$, $\Sigma_n$ is determined by the ACVF of an ARFIMA$(0,d,0)$ process, which has the exact form:
\begin{equation}
    \gamma_0(h) = \sigma^2 \frac{\Gamma(1-2d)}{\Gamma(1-d)\Gamma(d)}\frac{\Gamma(h+d)}{\Gamma(1+h-d)}. \label{eq:arfima0d0_acf}
\end{equation}

For $\pi(\phi,\theta |\bm{u}_n, d)$, $\Sigma_n$ is determined by the ACVF of an ARMA$(1,1)$ process, which has the exact form:
\begin{align*}
    \gamma(0) &= \sigma^2 \frac{1+2\phi\theta + \theta^2}{1-\phi^2}, \\
    \gamma(h) &= \phi \gamma(h-1), \text{ for } |h| > 1.
\end{align*}
Similar to the simultaneous MCMC of Section \ref{subsub:simultaneousMH}, the algorithm remains the same for \\ARFIMA$(1,d,0)$ and ARFIMA$(0,d,1)$ models, with $\theta$ or $\phi$ dropped respectively, and the relevant ARFIMA ACVF calculated with $\theta$ or $\phi$ set to zero. Since there are now two MH steps for the simulation of $d$ and $(\phi, \theta)$, the algorithm requires the specification of two proposal distributions labelled $p_1(\cdot)$ and $p_2(\cdot)$ respectively.
The parameter $\sigma^2$ has the same full conditional of Section \ref{subsub:simultaneousMH}. The full steps for the procedure are detailed in Algorithm \ref{algo:mcmc_method2}.

A key implication of this scheme compared to Algorithm \ref{algo:mcmc_method1} is the increased computational demand. While the direct computation of the ACVFs is simpler in this algorithm when compared to the required ARFIMA$(1,d,1)$ ACVF in Algorithm \ref{algo:mcmc_method1}, most of the computing time is spent in the inversions required in calculating the posterior distribution. This arises from filtering the series twice per simulation and performing an additional posterior calculation, which involves large $n\times n$ matrix operations. 

\begin{algorithm}[h]
\caption{ARFIMA$(1,d,1)$ Filtered MCMC } \label{algo:mcmc_method2}
Given series $\bm{y}_n$, hyperparameters $\alpha, \beta$, long-memory proposal $p_1(\cdot)$, ARMA proposal $p_2(\cdot)$, and initial values $(d_0, \phi_0, \theta_0)$,
\begin{algorithmic}
\For{iteration $i=1,\dots,M$}
    \medskip
    \State \textbf{Long-memory MH Random Walk Step}:
    \State Draw a proposal value $\tilde{d} \sim p_1(\cdot | d_{i-1})$.
    \State Filter $\bm{y}_n$ into an ARFIMA$(0,d,0)$ series $\bm{z}_n$ using $z_t = \phi_{i-1}(B)/\theta_{i-1}(B) y_t$.
    \State Calculate the acceptance probability for $\tilde{d}$
        \begin{align}
            \alpha_d = \min\Bigg\{0, \frac{\pi(\tilde{d} | \bm{z}_n, \phi_{i-1},\theta_{i-1},)}{\pi(d_{i-1}| \bm{z}_n, \phi_{t-1},\theta_{t-1})} \cdot \frac{p_1(d_{i-1}| \tilde{d})}{ p_1(\tilde{d}| d_{i-1})}\Bigg\}.
        \end{align}
    \State Simulate $w \sim \text{Bernoulli}(\alpha_d)$:
    \If{$w = 1$}
        \State Set $d_i = \tilde{d}$,
    \Else
        \State Set $d_i = d_{i-1}$.
    \EndIf
    \medskip
     \State \textbf{ARMA MH Random Walk Step}: 
     \State Draw proposal values $(\tilde{\phi},\tilde{\theta}) \sim p_2(\cdot |  \phi_{i-1},\theta_{i-1})$.
     \State Filter $\bm{y}_n$ into an ARMA$(1,1)$ series using $u_t= (1-B)^{d_i}y_t$.
    \State Calculate the acceptance probability for $(\tilde{\phi},\tilde{\theta})$
        \begin{align}
            \alpha_{\text{ARMA}} = \min\Bigg\{0, \frac{\pi(\tilde{\phi},\tilde{\theta} | \bm{u}_n, d_i)}{\pi(\phi_{i-1},\theta_{i-1} | \bm{u}_n, d_i)} \cdot \frac{p_2(\phi_{i-1},\theta_{i-1} | \tilde{\phi},\tilde{\theta})}{p_2(\tilde{\phi},\tilde{\theta} | \phi_{i-1},\theta_{i-1})}\Bigg\}.
        \end{align}  
    \State Simulate $w \sim \text{Bernoulli}(\alpha_\text{ARMA})$:
    \If{$w = 1$}
        \State Set $(\phi_i,\theta_i)= (\tilde{\phi},\tilde{\theta})$,
    \Else
        \State Set $(\phi_i,\theta_i)= (\phi_{i-1},\theta_{i-1})$.
    \EndIf

     \medskip
     \State \textbf{GS Step}: 
     \State Calculate the ARFIMA$(1,d,1)$ covariance matrix  $\Sigma_{n,i}$ using $(d_{i},\phi_i,\theta_i)$.
    \State Draw $\sigma_i^2 | d_{i}, \phi_i,\theta_i, \bm{y}_n \sim \text{IG}\big(\frac{n}{2} + \alpha,\frac{1}{2}(\bm{y}_n^T\Sigma_{n,i}^{-1} \bm{y}_n + 2\beta) \big)$.
\EndFor

\medskip
\State \textbf{Output}:
\State A set of MCMC samples $\{(d_i,\phi_i, \theta_i,\sigma^2_i) : i=1,\dots,M\}$ from $\pi(d,\phi,\theta,\sigma^2|\bm{y}_n)$.

\end{algorithmic}
\end{algorithm}

\subsection{ABC Framework for ARFIMA Models}
\label{sub:abc}

We now describe a novel ABC algorithm for approximating the posterior distribution $\pi(d,\phi,\theta,\sigma^2|\bm{y}_n)$. The MCMC implementation can be very slow, especially when $n$ is large. Our goal is to leverage the ease of parallelization inherent in the ABC algorithm to expedite the sampling from the posterior distribution, acknowledging that this comes at the cost of an increased approximation error. The algorithm we propose is a rejection sampling ABC algorithm. For its implementation, we need to specify proposal distributions for $(d,\phi,\theta,\sigma^2)$, define a set of lower-dimensional summary statistics $S(\cdot)$, and tolerance level $\epsilon$. As proposal distributions, we use the prior distributions as defined in Section \ref{sub:algo_for_bayes}. We discuss now how to select summary statistics and tolerance level.

When defining a summary statistic for a long-memory time series, the periodogram estimator of the spectral density function is an appealing choice. 
If a series exhibits long-memory, the low-frequency values of the spectral density should have a significantly larger magnitude than those in a series with short-memory \citep{leschinski2018periodogram}. The magnitude of these low-frequency values also scales with the value of $d$, implying that similar low-frequency values of the periodogram should reflect the strength of long-memory within a series. This feature suggests that the periodogram can be used as summary statistics, and it is possible to select a small number of periodogram points to reduce the dimension of the summary statistics. 

Determining the optimal number of periodogram points to compare is challenging. While low-frequency values capture substantial information about the long-memory behavior, there is a risk of information loss if only a portion of the periodogram values is considered. Conversely, comparing the full periodogram may include unnecessary higher frequency values that do not describe long-memory behavior, potentially leading to the curse of dimensionality. 

The periodogram is defined as
\begin{align}
    \mathcal{I}(k_j) &= \frac{1}{n} \bigg|\sum_{t=1}^{n} y_t e^{-it k_j} \bigg|^2,
    \label{q2:periodogram}
\end{align}
where it is evaluated at the Fourier frequencies $\bm{k}=\{k_1,\dots,k_n\}$, with $k_j = \frac{2\pi \cdot (j-1)}{n}$. Due to the symmetry of the periodogram, only $\frac{n}{2}$ values are needed (assuming $n$ is even), so $j=1, \ldots,n/2$. While the spectral density contains the same information as the series $\bm{y}_n$, the discrete periodogram is not a consistent estimator of the spectral density \citep{RN85}. Therefore, using the periodogram as a summary statistic inevitably involves some loss of information.

We consider three possible summary statistics related to the observed periodogram in the ABC rejection algorithm and denote the distance between the observed and simulated summary statistics as $H^{(h)}$ for $h=1,2,3$.

For an ABC iteration, given observed data $\bm{y}_n$ and simulated data $\bm{y}_\text{sim}$ of the same length $n$, the first variant uses the Euclidean distance between the complete periodograms:
\begin{align}
    H^{(1)} &= \bigg[\sum_{j=1}^{n/2}(\mathcal{I}_\text{obs}(k_j) - \mathcal{I}_\text{sim}(k_j))^2 \bigg]^{1/2},\notag
\end{align}
where $\mathcal{I}_\text{obs}$ and $\mathcal{I}_\text{sim}$ are the periodograms from the observed and simulated datasets, respectively.

The second distance uses only the first 20 values of the periodogram as summary statistics:
\begin{align}
    H^{(2)} &= \bigg[\sum_{j=1}^{20}(\mathcal{I}_\text{obs}(k_j) - \mathcal{I}_\text{sim}(k_j)^2 \bigg]^{1/2}.\notag
\end{align}
The choice of 20 values was determined through a separate study (results not shown here) that aimed to minimize information loss when drawing posterior samples using different subsets of the periodogram, and 20 was identified as the optimal choice. Since long-memory is indicated by high spectral density values at low frequencies, reducing the dimensionality of the summary statistic by including only the initial subset of the periodogram may still capture sufficient information for the ABC estimation.

Finally, we use the response variable from the log-periodogram regression model proposed by \cite{RN32} as summary statistic. This model has been shown to provide a consistent and asymptotically normally distributed OLS estimate, as well as a consistent Bayesian estimate for the parameter $d$, as demonstrated by \cite{grazian2024bayesian}. The response variable can be expressed as
\begin{small}
\begin{align}
    \mathcal{P}_i = \log \bigg\{ \sum_{j=1}^{J} \mathcal{I}(k_{i+j-J})  \bigg\}, \quad i\in I = \{\ell+J,\ell+2J,\dots,m\},\label{logperiodogram}
\end{align}
\end{small}
where $m$ is the number of frequencies to include, $J > 1$, and $\ell \geq 0$. 
This modified log-periodogram `pools' frequencies determined by the factor $J$, starting from a pre-determined trimming frequency $\ell$. For $J=1$ and $\ell=0$, $\mathcal{P}_i$ corresponds to the logarithm of the standard periodogram in Equation \eqref{q2:periodogram}. Similar to \texttt{R} package \texttt{LongMemoryTS} which implements the OLS estimator, in this work we set $J=2$ and $\ell=0$, employing the full range of frequencies with $m=n/2$ \citep{MAN_R3}. Therefore, the third distance metric is given by
\begin{align*}
    H^{(3)} &= \bigg[\sum_{i\in I} (\mathcal{P}_{\text{obs},i} - \mathcal{P}_{\text{sim},i})^2 \bigg]^{1/2},
\end{align*}
where $\mathcal{P}_{\text{obs},i}$ and $\mathcal{P}_{\text{sim},i}$ are the $i$-th log-periodogram regressor values for the observed and simulated datasets, respectively.

Finally, we also implemented a version of the ABC rejection algorithm using the OLS estimate for $d$ proposed by \cite{RN32}. The results of this version consistently showed a worse performance than the other versions, and are thus omitted. 

For parameters $(\phi,\theta)$, the ARFIMA maximum likelihood estimates (MLEs) are used as summary statistics. The distance metric $H^{(\text{ARMA})}$ is defined as the Euclidean distance between the MLEs of the observed and simulated series:
\begin{small}
\begin{align}
        H^{(\text{ARMA})} &= \bigg[ (\hat{\phi}_\text{MLE} - \hat{\phi}_\text{sim})^2 + (\hat{\theta}_\text{MLE} - \hat{\theta}_\text{sim})^2  \bigg]^{1/2},\notag
\end{align}
\end{small}
where $(\hat{\phi}_\text{MLE},\hat{\theta}_\text{MLE})$ are the ARFIMA$(1,d,1)$ MLE estimates for the observed series $\bm{y}_n$, and $(\hat{\phi}_\text{sim}, \hat{\theta}_\text{sim})$ are the ARFIMA$(1,d,1)$ MLE estimates for the simulated series $\bm{y}_{\text{sim}}$. Alternatively, the MLE values computed on the observed series and the simulated series can be obtained by filtering the series into an ARMA$(1,1)$ series using the simulated $\tilde{d}$, and represents the MLE values of the filtered series. Both methods have been implemented, returning similar results.

Finally, for $\sigma^2$, the distance metric $H^{(\sigma^2)}$ is the Euclidean distance between the sample variances of $\bm{y}_n$ and the simulated $\bm{y}_\text{sim}$:
\begin{equation*}
        H^{(\sigma^2)} = [(\Var(\bm{y}_n) - \Var(\bm{y}_{\text{sim}}))^2]^{1/2}.
\end{equation*}

The tolerance levels for accepting posterior draws are specified using two quantile values: one for the set of parameters $(d,\phi,\theta)$ and another for $\sigma^2$, denoted $q$ and $q_{\sigma^2}$ respectively. Acceptance is determined at the completion of the ABC simulations by sorting the saved distance metric vectors $(\bm{H}^{(h)}, \bm{H}^{(\text{ARMA})}, \bm{H}^{(\sigma^2)})$ for $h\in\{1,2,3\}$ and calculating the $q$-quantile values $(\epsilon_h,\epsilon_\text{ARMA})$ and $q_{\sigma^2}$-quantile value $(\epsilon_{\sigma^2})$ of the chosen distance metric. All posterior draws with distance metric $(\bm{H}^{(h)}, \bm{H}^{(\text{ARMA})}, \bm{H}^{(\sigma^2)})$ that simultaneously fall below each of their respective quantile thresholds are accepted. Since draws are only accepted if all their distance metrics fall below their thresholds, the number of accepted posterior draws will vary between runs. Full steps for the ABC procedure can be found in Algorithm \ref{algo:abc_sim}.

\begin{algorithm}[h]
\caption{ARFIMA$(1,d,1)$ ABC Rejection Sampling } \label{algo:abc_sim}
Given series $\bm{y}_{n}$, hyperparameters $\alpha, \beta$, distance function and summary statistics $H^{(h)},h\in\{1,2,3\}$, target distance quantiles $q$ and $q_{\sigma^2}$,
\begin{algorithmic}
\State Calculate the ARFIMA$(1,d,1)$ MLE for $(\phi,\theta)$ of $\bm{y}_n$, labelled $(\hat{\phi}_\text{MLE},\hat{\theta}_\text{MLE})$.
\State If $H^{(1)}$ or $H^{(2)}$ are used, calculate  $\mathcal{I}_{\text{obs}}$; if $H^{(3)}$ is used, calculate $\mathcal{P}_{\text{obs}}$.  
\medskip
\For{iteration $i=1,\dots,M$}
    \medskip
    \State \textbf{Simulation step}: 
    \State Simulate proposals $\tilde{d}_i, \tilde{\phi}_i, \tilde{\theta}_i,\tilde{\sigma}^2_i$ from their respective priors:
        \begin{align}
            \tilde{d}_i &\sim \text{Unif}(-0.5,0.5),\notag\\
            \tilde{\phi}_i &\sim \text{Unif}(-1,1),\notag\\
            \tilde{\theta}_i &\sim \text{Unif}(-1,1),\notag\\
            \tilde{\sigma}_i^2 &\sim \text{IG}(\alpha,\beta).\notag
        \end{align}
    \State Simulate a proposal series $\bm{y}_{\text{sim}} \sim \text{ ARFIMA}(1,d,1)$ of length $n$ using $(\tilde{d}_i,\tilde{\phi}_i, \tilde{\theta}_i,\tilde{\sigma}_i^2)$.

    \medskip
    \State \textbf{Summary statistics step}: 
    \State If $H^{(1)}$ or $H^{(2)}$ are used, calculate $\mathcal{I}_{\text{sim}}$ from the simulated $\bm{y}_{\text{sim}}$.
    \State If $H^{(3)}$ is used, calculate $\mathcal{P}_{\text{sim}}$ from the simulated $\bm{y}_{\text{sim}}$. 

    \medskip
    \State \textbf{Distance step}: 
    \State Calculate $H^{(h)}_i$, $H^{(\text{ARMA})}_{i}$ and $ H^{(\sigma^2)}_{i}$, representing the distance metric at iteration $i$.
    \State Store values $\tilde{d}_i,\tilde{\phi}_i, \tilde{\theta}_i,\tilde{\sigma}^2_i,H_i^{(h)},H^{(\text{ARMA})}_{i}, H^{(\sigma^2)}_{i}$. 
\EndFor

\medskip
\State \textbf{Acceptance step}: 
\State Calculate the tolerance level $\epsilon_h$ corresponding to the $q$-quantile of $\bm{H}^{(h)} = (H^{(h)}_1, \ldots, H^{(h)}_M)$.  
\State Calculate the tolerance level $\epsilon_\text{ARMA}$ corresponding to the $q$-quantile of $\bm{H}^{(\text{ARMA})} = (H^{(\text{ARMA})}_1,\ldots, H^{(\text{ARMA})}_M)$. 
\State Calculate the tolerance level $\epsilon_{\sigma^2}$ corresponding to the $q_{\sigma^2}$-quantile of $\bm{H}^{(\sigma^2)} = (H^{(\sigma^2)}_1, \ldots, H^{(\sigma^2)}_M)$. 
\State Accept simulated values $(\tilde{d}_i,\tilde{\phi}_i, \tilde{\theta}_i,\tilde{\sigma}^2_i)$ if their respective associated distance metrics $H_i^{(h)}$, $H^{(\sigma^2)}_i$ and $H_i^{(\text{ARMA})}$ \textit{simultaneously} fall below the $q$ and $q_{\sigma^2}$-quantile thresholds $(\epsilon_h,\epsilon_\text{ARMA},\epsilon_{\sigma^2})$. 

\medskip
\State \textbf{Output}:
\State A set of accepted ABC samples $\{(\tilde{d}_i,\tilde{\phi}_i, \tilde{\theta}_i,\tilde{\sigma}^2_i) : i=1,\dots,\tilde{M}\}$ from the approximate $\pi(d,\phi,\theta,\sigma^2 | \bm{y}_n)$, where $\tilde{M} \leq M$ is the number of accepted samples. 
 
\end{algorithmic}
\end{algorithm}

\subsection{ARFIMA  Forecasting} \label{sec_arfimaforecast}

A Bayesian approach to time series modeling offers a natural solution to the problem of forecasting, with the posterior predictive density capturing uncertainty regarding forecasts.

For an ARFIMA series $\bm{y}_n$, the posterior predictive density represents the density of an unobserved part of the series $\tilde{\bm{y}}_b$ of length $b$, given the observed series:
\begin{align}
    p(\tilde{\bm{y}}_b | \bm{y}_n) = \int_{\Psi} p(\tilde{\bm{y}}_b | \bm{y}_n, \psi) p(\psi | \bm{y}_n) d\psi. \notag
\end{align}

For an ARFIMA model, new series points depend on previously observed points, meaning that $p(\tilde{\bm{y}}_b | \bm{y}_n, \psi)$ must be evaluated directly due to the dependency on $\bm{y}_n$. In order to reduce the computational cost, \cite{RN3} express the density as a product of the conditional distributions of each series point using partial linear regression coefficients. Here, we implement a different approach. 

A more direct but computationally demanding method for forecasting requires evaluating the density directly. Under the assumption of Gaussian realizations, the conditional density can be derived using standard marginal multivariate normal results. Define $\bm{y}_{n+b} = (\bm{y}_n, \tilde{\bm{y}}_b)$, then 
\begin{align}
    \bm{y}_{n+b} | \psi \sim N_{n+b}(\bm{0}, \sigma^2 \Sigma_{n+b}),\notag 
\end{align}
where $\Sigma_{n+b}$ is the covariance matrix consisting of the ARFIMA autocorrelations for $n+b$ lags. Partitioning the covariance matrix into sections:
\begin{align}
\Sigma_{n+b} &= \begin{pmatrix}
\Sigma_n & \Sigma_{12} \\
\Sigma_{21} & \Sigma_{22}
\end{pmatrix}, \notag
\end{align}
the predictive density given previously observed $\bm{y}_n$ is derived as
\begin{align}
    \tilde{\bm{y}}_b | \bm{y}_n, &\psi \sim N_b(\bar{\bm{\mu}}, \bar{\Sigma}_b),
    \label{post_predictive}
\end{align}
with:
\begin{align}
    \bar{\bm{\mu}} &= \Sigma_{21} \cdot \Sigma_n^{-1} \cdot \bm{y}_n, \notag\\
    \bar{\Sigma}_b &= \Sigma_{22} - \Sigma_{21} \cdot \Sigma_n^{-1} \cdot \Sigma_{12}.\notag
\end{align}

Draws from $p(\tilde{\bm{y}}_b | \bm{y}_n)$ can be obtained at each iteration of the MCMC Algorithms \ref{algo:mcmc_method1} or \ref{algo:mcmc_method2}, or for each set of accepted parameters after running the ABC Algorithm \ref{algo:abc_sim}. 

\section{Simulation Study}
\label{section4}

\subsection{Simulation setting}

The MCMC and ABC methods presented in Section \ref{section3} were implemented for ARFIMA$(1,d,1)$, ARFIMA$(1,d,0)$ and ARFIMA$(0,d,1)$ models in an extensive simulation study. This study used several combinations of parameters $(d,\phi,\theta)$, as shown in Table \ref{table:par_combinations}, with $\sigma^2$ fixed at 1.

Each parameter set in Table \ref{table:par_combinations} was used to generate 200 simulated ARFIMA series, each of length $n=1,000$. 

\begin{table*}[h]
\centering
\caption{Simulation Study Parameter Combinations}\label{table:par_combinations}
\begin{tabular}{ |c|c|c|c||c|c|c||c|c|c| } 
 \hline
 Model & $d$ & $\phi$ & $\theta$ & Model & $d$ & $\phi$ & Model & $d$ & $\theta$ \\ \hline
 ARFIMA$(1,d,1)$ & $0.05$ & $0.2$ & $0.2$ & ARFIMA$(1,d,0)$ & $0.05$ & $0.2$ & ARFIMA$(0,d,1)$ & $0.05$ & $0.2$\\
 & $0.05$ & $0.5$ & $0.5$ & & $0.05$ & $0.5$ & & $0.05$ & $0.5$ \\
 & $0.1$  & $0.2$ & $0.2$ & & $0.1$ & $0.2$ & & $0.1$ & $0.2$ \\
 & $0.1$  & $0.5$ & $0.5$ & & $0.1$ & $0.5$ & & $0.1$ & $0.5$ \\
 & $0.2$  & $0.2$ & $0.2$ & & $0.2$ & $0.2$ & & $0.2$ & $0.2$ \\
 & $0.2$  & $0.5$ & $0.5$ & & $0.2$ & $0.5$ & & $0.2$ & $0.5$ \\
 & $0.3$  & $0.2$ & $0.2$ & & $0.3$ & $0.2$ & & $0.3$ & $0.2$ \\
 & $0.3$  & $0.5$ & $0.5$ & & $0.3$ & $0.5$ & & $0.3$ & $0.5$ \\
 & $0.4$  & $0.2$ & $0.2$ & & $0.4$ & $0.2$ & & $0.4$ & $0.2$ \\
 & $0.4$  & $0.5$ & $0.5$ & & $0.4$ & $0.5$ & & $0.4$ & $0.5$\\
 \hline
\end{tabular}
\end{table*}

\subsubsection{Hyperparameters and Proposal Distributions}\label{subsub:hyperparam_proposal}

Prior distributions were selected as follows. Uniform independent prior distributions were chosen for $d \sim Unif(-0.5,0.5)$, $\phi \sim Unif (-1,1)$, and $\theta \sim Unif(-1,1)$. The prior for $\sigma^2$ was set as an informative inverse gamma distribution $IG(\alpha, \beta)$. The hyperparameters of the $\sigma^2$ prior distribution were set to  $\alpha=28$ and $\beta=30$, resulting in a density centered around 1, with a 95\% interval of [0.76, 1.61]. This choice was made because the acceptance rates of the ABC algorithms can be influenced by the use of noninformative prior distributions when used as proposal distributions. Simulations were also repeated using a less informative $IG(1,1)$ prior distribution with 95\% interval of [0.27,39.50], yielding MCMC results consistent with the findings presented here. 

For the simultaneous MCMC method in Algorithm \ref{algo:mcmc_method1}, the proposal distribution for $(\tilde{d},\tilde{\phi},\tilde{\theta})$ at iteration $i=1,\dots,M$ is a truncated multivariate normal distribution $N_{tr}([d_{i-1},\phi_{i-1},\theta_{i-1}], \Sigma_{\text{MLE,s}})$ with bounds $d \in (-0.5,0.5)$, $\phi \in (-1,1)$, and $\theta \in (-1,1)$. Here, $[d_{i-1},\phi_{i-1},\theta_{i-1}]$ represents parameter values chosen at the previous iteration, and $\Sigma_{\text{MLE,s}}$ is the ARFIMA MLE covariance matrix. The initial values for $(d_0, \phi_0, \theta_0)$ are sampled from the same truncated multivariate normal distribution, with the mean parameter set to the ARFIMA MLE estimates of $(d, \phi, \theta)$.

For the filtered MCMC method in Algorithm \ref{algo:mcmc_method2}, two proposal distributions are required for $\tilde{d}$ and $(\tilde{\phi}, \tilde{\theta})$, labeled $p_1(\cdot)$ and $p_2(\cdot)$, respectively. For iteration $i=1,\dots,M$, the first proposal is a truncated normal distribution $N_{tr}(d_{i-1}, \sigma^2_d)$ with bounds $d\in(-0.5,0.5)$, where $d_{i-1}$ is the chosen parameter at the previous iteration and $\sigma^2_d$ is the hyperparameter proposal variance. The hyperparameter value was set to $\sigma^2_d = 0.025^2$, chosen such that $95\%$ interval covers approximately a 0.1 range around $d_{i-1}$.
The second proposal for the ARMA parameters is a truncated multivariate normal distribution $N_{tr}([\phi_{i-1},\theta_{i-1}], \Sigma_{\text{MLE,f}})$ with bounds $\phi\in (-1,1)$, and $\theta\in (-1,1)$, where $\Sigma_{\text{MLE,f}}$ is the ARMA MLE covariance matrix, calculated from the series $u_t = (1-B)^{d_0}y_t$. The initial value $d_0$ is drawn from its proposal distribution, with the mean set to the ARFIMA MLE estimate for $d$. After $d_0$ is simulated, $(\phi_0, \theta_0)$ are simulated from their proposal distribution, with the mean set to the ARMA MLE estimates of $(\phi,\theta)$ calculated from the series $\bm{u}_t$.  

For the ABC simulation, three values of the acceptance quantile $q$, used to determine the long-memory and ARMA tolerance levels $\epsilon_h$ and $\epsilon_\text{ARMA}$, respectively, were compared, with values $q = \{0.01,0.005,0.001\}$. The acceptance quantile for $\sigma^2$ was set to $q_{\sigma^2} = 0.5$. 

\subsubsection{Computation of the ACVF}

Algorithm \ref{algo:mcmc_method1} and \ref{algo:mcmc_method2} both require the calculation of the exact autocovariance function (ACVF) of an ARFIMA$(1,d,1)$ process. However, evaluating the hypergeometric function in \eqref{eq:Cfun} is computationally demanding. Additionally, there is a scalability issue for higher orders $p,q > 1$, where determining the exact form of the ACVF becomes challenging. To improve the speed of the MCMC methods, we use a convolution-based approach between the ACVF of the long-memory component and the ACFV of the ARMA components, utilizing the \texttt{R} package \texttt{arfima} \citep{RN40}. When compared against exact ACVF values across a grid of $(d,\phi,\theta)$ values, this method showed negligible differences while offering significant speed improvements. In Algorithm \ref{algo:mcmc_method2}, the exact forms of the ACVF for ARFIMA$(0,d,0)$ and ARMA$(1,1)$ processes are still used due to being computationally simple.

\subsubsection{Run Details}

The following Bayesian methods were run, and we indicate here their respective abbreviations:
\begin{enumerate}
    \item MCMC: MCMC algorithm of \cite{RN3}, as described in Algorithm \ref{algo:mcmc_method1}; 
    \item MCMC Filter: Filtered MCMC as described in Algorithm \ref{algo:mcmc_method2};
    \item ABC FP-Q$\nu$: ABC simulation as described in Algorithm \ref{algo:abc_sim} using the full-periodogram as the summary statistic and distance $H^{(1)}$, with samples accepted with $(\epsilon_h, \epsilon_{\text{ARMA}})$ determined by $q=\nu$;
    \item ABC 20P-Q$\nu$: ABC simulation as described in Algorithm \ref{algo:abc_sim} using 20 values of the periodogram as the summary statistic and distance $H^{(2)}$, with samples accepted with $(\epsilon_h, \epsilon_{\text{ARMA}})$ determined by $q=\nu$;
    \item ABC LogP-Q$\nu$: ABC simulation as described in Algorithm \ref{algo:abc_sim} using the log-periodogram regression model response variable as the summary statistic and distance $H^{(3)}$, with samples accepted with $(\epsilon_h, \epsilon_{\text{ARMA}})$ determined by $q=\nu$.
\end{enumerate}

The results of the simulation study were summarised using the following metrics:
\begin{enumerate}
    \item Mean: Average posterior mean computed across  repetitions;
    \item LB: Average lower bound for the 90\% Bayesian credible interval computed across repetitions;
    \item UB: Average upper bound for the 90\% Bayesian credible interval computed across repetitions;
    \item Coverage: Proportion of repetitions where the true parameters values are included in the 90\% Bayesian credible interval;
    \item SD: Average standard deviation computed across repetitions;
    \item RMSE: Root mean squared error computed across repetitions;
    \item ESS$/ \bar{m}$: Either the Effective Sample Size (ESS) for MCMC methods or the average number $\bar{m}$ of accepted values across repetitions for ABC methods. Note that $\bar{m}$ is not constant for quantile $q$ since parameters $(d,\phi,\theta,\sigma^2)$ are accepted only when they simultaneously fall below their respective tolerance levels.
\end{enumerate}

For the MCMC chains, we observed high serial correlation between the posterior draws, particularly in the MCMC Filter method. This correlation was especially strong for the $d$ and $\phi$ chains, but less pronounced for $\theta$ and $\sigma^2$. To address this, thinning was applied by retaining one in every 50 draws and discarding the rest for both MCMC and MCMC Filter. From the $M=50,000$ MCMC simulations, $1,000$ posterior draws were retained after thinning. ACF plots were analyzed post-thinning, showing autocorrelations close to zero beyond lag 10.
For the ABC runs, $M=15$ million simulations were performed.

In addition to the Bayesian methods, frequentist estimation results are also included for each repetition, using the ARFIMA MLE estimates of the relevant parameters. For the frequentist results, the 90\% confidence intervals were computed. MLE estimates are calculated using the \texttt{R} package \texttt{nsarfima} \citep{MAN_R1}.

Within each repetition of the simulation study, the sampled posterior parameters were used to forecast the next 15 values using the method outlined in Section \ref{sec_arfimaforecast}. Since the data for the study is stochastically generated for each repetition, the `true' next 15 points can be calculated by extending the target series to $b = n + 15 = 1,015$ points. The posterior parameters are simulated on the first $n = 1,000$ points, while the `future' $15$ points are used for forecast validation. For ABC methods, forecasting results were only generated for the $(d,\phi,\theta)$ acceptance quantile $q=0.01$. Acceptance quantiles $q = 0.005$ and $q=0.001$ were omitted due to the small number of accepted parameters for certain configurations, with some acceptance numbers being less than 500. For the sampled posterior of both MCMC and ABC methods, one forecasting series of $15$ points was generated for each accepted posterior draw. 

For the 200 repetitions for a given set of parameters, the forecasting RMSE, SD, and 90\% Bayesian coverage intervals were calculated. The coverage is computed as the average coverage across all 15 points and all repetitions for each parameter set. Frequentist forecast predictions were also calculated with the \texttt{R} package \texttt{forecast}, using MLE estimates \citep{MAN_R2}.

\subsection{Inferential Results}

Simulation output tables can be found in Appendices \ref{appendix:arfima1d1}, \ref{appendix:arfima1d0}, and \ref{appendix:arfima0d1}. Although the quantile $q=0.001$ was included in the ABC simulations, the corresponding results are omitted here due to too few values being accepted for a valid comparison. There is some variation in the best-performing model depending on the magnitude of $(d, \phi, \theta)$; however, in general, among the Bayesian methods, MCMC Filter performed best in terms of minimizing SD, bias, and RMSE, followed by MCMC, ABC 20P, and ABC FP. ABC 20P showed similar results to ABC FP, with performance metrics being close between the two methodologies. The ABC LogP method tended to perform the worst, generally exhibiting the largest SD and widest credible intervals. Appendix \ref{appendix:bestperforming} presents a count of the number of times each method achieved the lowest bias, RMSE, SD, and highest coverage. Note that some columns sum to greater than 10 due to ties in the metrics. For the ABC models, there is little difference in results between the $0.01$ and $0.005$ quantiles, except for the reduced number of accepted samples. 

In general, the posterior densities from both ABC and MCMC Bayesian methods captured the presence and strength of long-memory through the estimation of $d$. Notably, in all cases, including frequentist results, the posterior mean or the maximum likelihood estimator for $d$ underestimated the true parameter. This downward bias is most prominent in ABC methodologies and is less pronounced in MCMC and frequentist results. The performance of each method in estimating $d$ varied across different ARFIMA configurations. A particularly significant trend was the increased range of the credible interval in the presence of a large AR parameter $\phi=0.5$. For ARFIMA$(1,d,1)$ and ARFIMA$(1,d,0)$ models with small $d<0.2$ and large $\phi=0.5$, credible interval ranges often extended into negative $d < 0$.

When comparing Bayesian results to frequentist results, MLE generally exhibited lower bias on average, with estimates closer to the true long-memory parameter. However, the RMSE tended to be lower for MCMC methods (and sometimes for ABC methods). While the SD results were lower for the frequentist fit, these asymptotic values did not always capture the uncertainty surrounding the target parameters as effectively as the Bayesian methods. Frequentist results tended to poorly estimate $d$ in the presence of a large AR parameter $\phi=0.5$, with coverage values for $d$ dropping to 56-70\% in these cases. This phenomenon, related to the interaction between the AR component and the long-memory component, appears to be poorly accounted for in frequentist MLE estimation. When $\phi=0.2$, the estimation of $d$ improved significantly across all methods. For ARFIMA$(0,d,1)$ models, the magnitude of the MA parameter $\theta$ had relatively little impact on the estimation of $d$ across all methods, yielding results with the best credible interval, SD, and RMSE metrics.

In estimating $d$, the MCMC Filter method appears to offer improvements over the standard MCMC method. For almost all model configurations, the SD, RMSE, and coverage are improved by filtering the ARFIMA chain. In the ABC FP and ABC 20P methods, the inclusion of only the first 20 periodogram values tended to slightly lower the SD and RMSE, though this came at the cost of marginally increasing the coverage. ABC LogP methods often exhibited very large coverage.

Credible intervals were typically wider for $d=0.05$; the ABC methods, in particular, had lower bounds that extended into negative values. This seemed to be due to a small number of accepted simulations yielding negative values approaching $d=-0.5$, and can be potentially solved by increasing the number of simulations while keeping the tolerance level fixed.

ESS values varied depending on the method and parameter values. For $\theta$ and $\sigma^2$, ESS values were typically greater than 850, while $d$ and $\phi$ ESS values varied with the true value of $\phi$, and to a lesser extent, the true value of $d$. The standard MCMC method produced larger ESS values for $(d,\phi)$ compared to MCMC Filter, reflecting the higher serial autocorrelation observed in the ACF plots. For ARFIMA$(1,d,1)$ models, the MCMC Filter ESS values for $(d,\phi)$ ranged between 650 and 400 for $\phi=0.2$ and $\phi=0.5$ respectively, while the MCMC ESS values approximately ranged between 750 and 500. For ARFIMA$(1,d,0)$ models, the MCMC Filter ESS values for $(d,\phi)$ ranged between 750 and 450 for $\phi=0.2$ and $\phi=0.5$, while the MCMC ESS values were around 800 and 950. ESS values for ARFIMA$(0,d,1)$ models were consistently around 1000.

The ARMA parameters $(\phi,\theta)$ were generally well-estimated by all Bayesian methods, even in the presence of large $d$ and large $\phi=0.5$. Coverage percentages typically ranged around 80-90\%. The MCMC Filter method, followed by MCMC and ABC FP methods, consistently captured the ARMA parameters with the shortest credible intervals. Frequentist results performed well in most cases; however, similar to the poor estimates of $d$ in the presence of large $\phi=0.5$, the coverage for $\phi$ was also suboptimal when the true parameter was $\phi=0.5$. Appendix \ref{appendix:arfima1d0} shows that, for an ARFIMA$(1,d,0)$ model, the frequentist coverage for $\phi$ is usually below the expected 90\% and can drop to 64\%, an issue not present in the Bayesian methods.

For Bayesian methods, the ability to accurately capture the true parameters $(\phi,\theta)$ generally increased with larger $(\phi, \theta)$, resulting in tighter credible intervals and smaller standard deviations when $\phi=\theta=0.5$ compared to $\phi=\theta=0.2$.

Among the Bayesian methods, the variance $\sigma^2$ is best captured by the MCMC Filter and MCMC methods, followed by ABC LogP and ABC 20P. The ABC FP method estimates $\sigma^2$ reasonably well for most parameter combinations, but it performs poorly when the long-memory parameter is small, at $d=0.05$. In these cases, the coverage for $\sigma^2$ often falls below 10\%, with the interval failing to capture the true $\sigma^2 = 1$. This issue was not observed in the ABC 20P results. Frequentist estimates of $\sigma^2$ generally exhibit a coverage higher than expected for ARFIMA(1,d,1) models with respect to Bayesian methods.

\subsection{Forecasting Results}

\begin{figure*}[h]
\begin{center}
\includegraphics[width=1.0\textwidth]{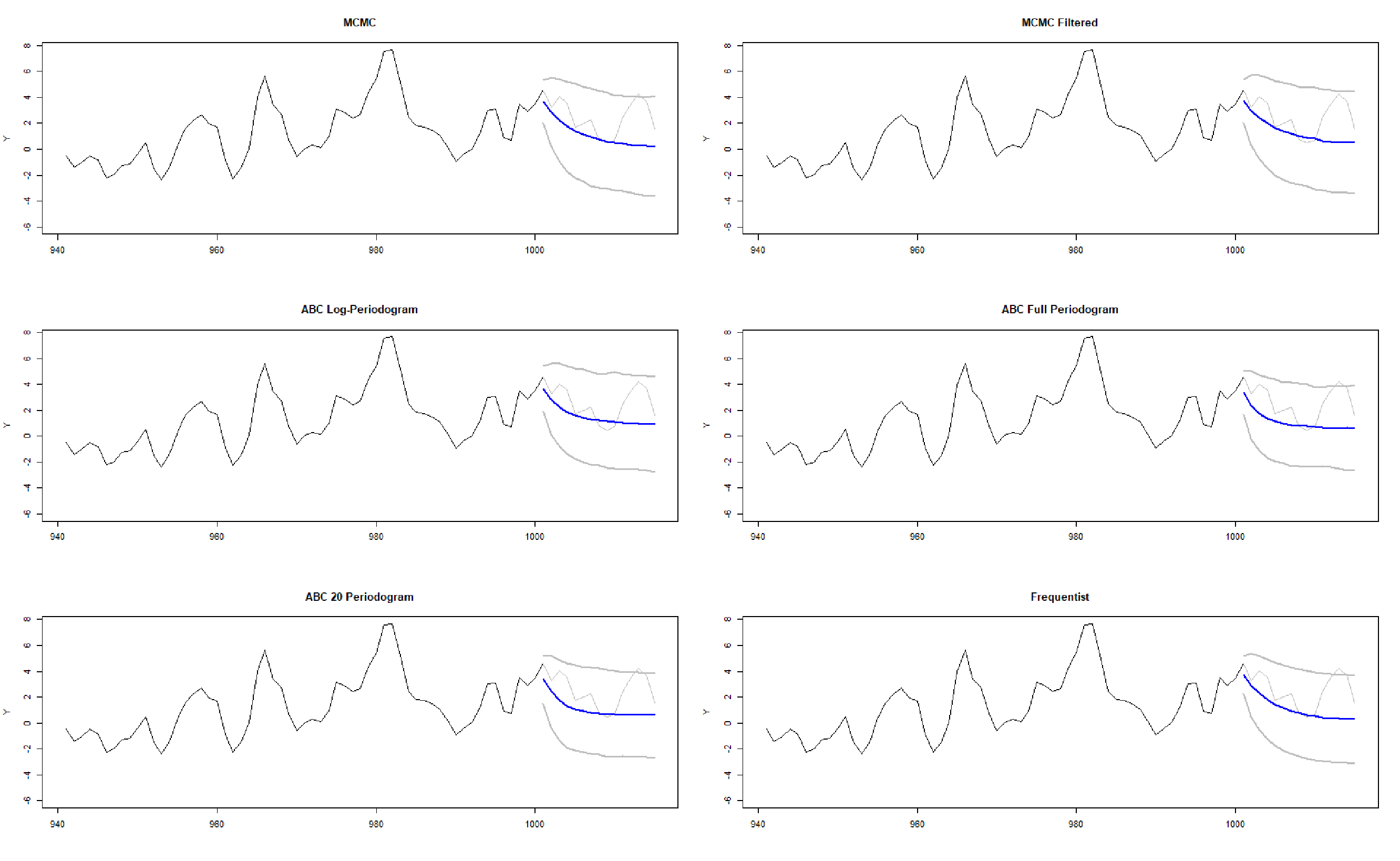}
\end{center}
\caption{Forecast for the next $15$ points of an ARFIMA$(1,d,1)$ series of length $n=1000$ with $d=0.3, \phi=0.5, \theta=0.5$. The true next $15$ points are coloured in grey} 
\label{plots:forecast_study_sample}
\end{figure*}

Forecasting results can be found in Appendix \ref{appendix:forecast_results}. Generally, across the ARFIMA$(1,d,1)$, ARFIMA$(1,d,0)$, and ARFIMA$(0,d,1)$,  models, results are consistent among the MCMC, ABC, and frequentist methodologies. The MCMC Filter method tends to perform best in terms of coverage, with MCMC results being comparable to frequentist results overall. ABC results typically exhibit slightly lower coverage; however, for ABC with the full periodogram, SD is notably lower for some model specifications compared to other models. For a 15-point forecast, the approximately 90\% coverage across methods corresponds to around 1-2 points being not included from the credible interval on average across repetitions. A sample forecasting plot is shown in Figure \ref{plots:forecast_study_sample}, demonstrating one simulated dataset with long-memory parameter of $d=0.3$ and ARMA parameters $\phi=\theta=0.5$.

Similar to the simulation results, RMSE and SD values tend to be larger when $\phi=0.5$ or $\theta = 0.5$. The same trend is observed for $d$, with larger long-memory parameter values corresponding to increased RMSE and SD. For all models, the greatest variance and error were found in runs with $d=0.4$, $\phi=0.5$, and/or $\theta=0.5$. Despite this, coverage tended to remain consistent across parameter combinations.

While forecasting ARFIMA models may be useful for capturing potential future trends, prediction intervals appear to be large across both Bayesian and frequentist methods, even for near-term points. 

\section{Application}
\label{section5}

We now provide an application of the Bayesian MCMC and ABC methods to a financial dataset containing seasonally adjusted quarterly GNP values in the United States from January 1, 1947, to October 1, 2023 \citep{RN75}. Several authors have studied the potential presence of long-memory within GNP, with some suggesting that long-memory models may describe the series better than short memory models \citep{RN71, RN73, RN74}. However, there is also some indication that long-memory may be due to non-normality or non-linearity \citep{RN72}. \cite{RN3} also applied their MCMC algorithm to GNP data. However, their analysis included fewer quarterly GNP records than are now available, with the ARFIMA model being fit on 172 data points up to the year 1991.

We consider the log-returns of the quarterly GNP price series $\bm{x}_n$ which yields $n=308$ data points in the series $\bm{y}_n$:
\begin{align}
    y_t &= (\log x_t - \log x_{t-1}) \cdot 100 \qquad t=1, \ldots, n. \notag
\end{align}
The data is then de-meaned by subtracting the series mean from each data point. We fit \\ ARFIMA$(1,d,1)$, ARFIMA$(1,d,0)$ and ARFIMA$(0,d,1)$ models using the algorithms presented in Section \ref{section3}.

The prior distributions followed the same form as the simulation study, with $\sigma^2$ hyperparameters set to $\alpha=33$ and $\beta=45$, giving a density with a $95\%$ interval of $[1,2]$. These parameters were chosen such that it contains the series sample variance of $1.683$. 

For the MCMC methods, $50,000$ iterations were calculated with a thinning rate of $50$, returning $1,000$ posterior draws. The correlation of the chains were verified through the ACF plots of each parameter, yielding insignificant autocorrelations after thinning at lag greater than $10$. The MCMC proposal distributions followed the same configuration as described in Section \ref{subsub:hyperparam_proposal}.

For the ABC methods, 15 million iterations were run using the same three summary statistics presented in Section \ref{sub:abc}. The acceptance quantile for the $d$ and $(\phi,\theta)$ parameters was set to $q=0.01$ and the acceptance quantile for $\sigma^2$ was set to $q_{\sigma^2}=0.5$.

The means and standard deviations for the different model fits are presented in Table \ref{gnp_arfima1d1_table}, Table \ref{gnp_arfima1d0_table}, and Table \ref{gnp_arfima0d1_table}, with the corresponding density plots available in Appendix \ref{appendix:gnp_model_densities}. Across ARFIMA$(1,d,1)$, ARFIMA$(1,d,0)$ and ARFIMA$(0,d,1)$ models, estimates for $d$ consistently ranged between $[0.2,0.25]$. The long-memory densities were slightly more concentrated for the MCMC and MCMC Filter fits compared to the ABC results. Among the ABC results, ABC 20P and ABC FP produced similar outcomes, with smaller standard deviations than ABC LogP. The posterior mean of $d$ was lower for the ABC methods compared to the MCMC methods, mirroring the findings from the simulation study. Results between the MCMC and MCMC Filter methods were similar, with no significant differences.

\begin{table}[h]
\centering
\caption{ARFIMA$(1,d,1)$ GNP Results}\label{gnp_arfima1d1_table}
\centering
\begin{tabular}[t]{l|l|r|r}
\hline
Par & Method & Mean & SD\\
\hline
d & MCMC & 0.22773 & 0.09695\\
\hline
$\phi$ & MCMC & -0.01173 & 0.56668\\
\hline
$\theta$ & MCMC & -0.02353 & 0.53625\\
\hline
$\sigma^2$ & MCMC & 1.53199 & 0.12249\\
\hline
d & MCMC Filter & 0.21701 & 0.08188\\
\hline
$\phi$ & MCMC Filter & -0.06110 & 0.44680\\
\hline
$\theta$ & MCMC Filter & 0.03279 & 0.39629\\
\hline
$\sigma^2$ & MCMC Filter & 1.53657 & 0.12705\\
\hline
d & ABC LogP-Q0.01 & 0.18135 & 0.19401\\
\hline
$\phi$ & ABC LogP-Q0.01 & -0.03342 & 0.56731\\
\hline
$\theta$ & ABC LogP-Q0.01 & -0.01231 & 0.48208\\
\hline
$\sigma^2$ & ABC LogP-Q0.01 & 1.26812 & 0.11499\\
\hline
d & ABC FP-Q0.01 & 0.17866 & 0.16376\\
\hline
$\phi$ & ABC FP-Q0.01 & -0.03206 & 0.55805\\
\hline
$\theta$ & ABC FP-Q0.01 & -0.02406 & 0.48735\\
\hline
$\sigma^2$ & ABC FP-Q0.01 & 1.15574 & 0.12849\\
\hline
d & ABC 20P-Q0.01 & 0.19985 & 0.13500\\
\hline
$\phi$ & ABC 20P-Q0.01 & -0.10517 & 0.53418\\
\hline
$\theta$ & ABC 20P-Q0.01 & 0.00476 & 0.49394\\
\hline
$\sigma^2$ & ABC 20P-Q0.01 & 1.30014 & 0.19217\\
\hline
\end{tabular}
\end{table}

\begin{table}[h]
\centering
\caption{ARFIMA$(1,d,0)$ GNP Results}\label{gnp_arfima1d0_table}
\centering
\begin{tabular}[t]{l|l|r|r}
\hline
Par & Method & Mean & SD\\
\hline
d & MCMC & 0.23249 & 0.07496\\
\hline
$\phi$ & MCMC & -0.04251 & 0.09408\\
\hline
$\sigma^2$ & MCMC & 1.53958 & 0.12476\\
\hline
d & MCMC Filter & 0.23443 & 0.07372\\
\hline
$\phi$ & MCMC Filter & -0.04187 & 0.09412\\
\hline
$\sigma^2$ & MCMC Filter & 1.54285 & 0.13021\\
\hline
d & ABC LogP-Q0.01 & 0.20669 & 0.12915\\
\hline
$\phi$ & ABC LogP-Q0.01 & -0.04819 & 0.09303\\
\hline
$\sigma^2$ & ABC LogP-Q0.01 & 1.26585 & 0.11244\\
\hline
d & ABC FP-Q0.01 & 0.19638 & 0.10489\\
\hline
$\phi$ & ABC FP-Q0.01 & -0.04917 & 0.09024\\
\hline
$\sigma^2$ & ABC FP-Q0.01 & 1.15815 & 0.12909\\
\hline
d & ABC 20P-Q0.01 & 0.19204 & 0.09421\\
\hline
$\phi$ & ABC 20P-Q0.01 & -0.05183 & 0.08447\\
\hline
$\sigma^2$ & ABC 20P-Q0.01 & 1.28427 & 0.18857\\
\hline
\end{tabular}
\end{table}

\begin{table}[h]
\centering
\caption{ARFIMA$(0,d,1)$ GNP Results}\label{gnp_arfima0d1_table}
\centering
\begin{tabular}[t]{l|l|r|r}
\hline
Par & Method & Mean & SD\\
\hline
d & MCMC & 0.22880 & 0.06693\\
\hline
$\theta$ & MCMC & -0.03531 & 0.08026\\
\hline
$\sigma^2$ & MCMC & 1.53651 & 0.11934\\
\hline
d & MCMC Filter & 0.23156 & 0.06729\\
\hline
$\theta$ & MCMC Filter & -0.03857 & 0.07918\\
\hline
$\sigma^2$ & MCMC Filter & 1.53959 & 0.12391\\
\hline
d & ABC LogP-Q0.01 & 0.20548 & 0.13050\\
\hline
$\theta$ & ABC LogP-Q0.01 & -0.04840 & 0.09523\\
\hline
$\sigma^2$ & ABC LogP-Q0.01 & 1.26693 & 0.11402\\
\hline
d & ABC FP-Q0.01 & 0.20798 & 0.10901\\
\hline
$\theta$ & ABC FP-Q0.01 & -0.04978 & 0.09921\\
\hline
$\sigma^2$ & ABC FP-Q0.01 & 1.15999 & 0.12922\\
\hline
d & ABC 20P-Q0.01 & 0.20731 & 0.10350\\
\hline
$\theta$ & ABC 20P-Q0.01 & -0.05266 & 0.09968\\
\hline
$\sigma^2$ & ABC 20P-Q0.01 & 1.26974 & 0.16926\\
\hline
\end{tabular}
\end{table}

Non-zero estimates for $d$ were also reported in \cite{RN3}, where several ARFIMA models of varying orders were fitted to 172 GNP data points using MCMC methods. Their results favored the ARFIMA$(0,d,0)$ model, yielding a long-memory estimate of $\hat{d}=0.36$. Although their paper also fitted ARFIMA$(1,d,1)$, ARFIMA$(1,d,0)$, and ARFIMA$(0,d,1)$models, these results were unfortunately omitted and cannot be compared to our findings. In a similar analysis, \cite{RN93} used maximum likelihood estimates to fit ARFIMA models on 168 GNP points. Their estimate for the long-memory parameter in the ARFIMA$(0,d,1)$ model was lower, at  $\hat{d}=0.16$, while the long-memory estimates for the ARFIMA$(1,d,1)$ and ARFIMA$(1,d,0)$ models were negative. However, comparing these results is challenging since our analysis uses almost twice as many quarterly GNP points. We also fitted an ARFIMA$(0,d,0)$ model using only the MCMC method with a thinning rate of one in 50, yielding a posterior mean of $\hat{d}=0.205$, which is closer to the estimate obtained with other models, and a standard deviation of $0.044$. The corresponding density plot is shown in Appendix \ref{plots:gnp_arfima0d0}.

Estimates for $\sigma^2$ are consistent across model fits but differ slightly in magnitude between the MCMC and ABC results. The MCMC and MCMC Filter methods returned posterior means of approximately $\hat{\sigma}^2=1.5$, while the ABC methods produced slightly lower values, with ABC FP yielding $\hat{\sigma}^2=1.15$ and ABC LogP and ABC 20P yielding $\hat{\sigma}^2=1.25$. Estimates for the ARMA parameters in the ARFIMA$(1,d,0)$ and ARFIMA$(0,d,1)$ models were similar between the MCMC and ABC methods, with results showing a tight density centered around zero.

ARFIMA$(1,d,1)$ results show bimodal posterior marginal densities for $\phi$ and $\theta$, which can be an indication of model misspecification. Repeated calculations of the MLEs showed a similar issue, with results alternating between similar positive and negative estimates. The mean MLE value from these repetitions for $d$ was consistent with the Bayesian results, yielding a value of $\hat{d}=0.22$ with a standard deviation of $0.087$.

To choose the best model fitting the observed data, we calculated the Bayesian deviance information criterion (DIC) \citep{RN92}. DIC values for the ARFIMA$(0,d,0)$ model were also included for comparison. However the DIC results showed insignificant differences between the models fits, not providing enough information to favour one specific model over another. The relevant DIC values for the MCMC method are shown in Table \ref{gnp_dic_table}; MCMC Filter DIC values were also calculated and confirmed to be similar to the MCMC method. The Bayes Factor was also computed, yielding similar values across models.

\begin{table}[h]
\centering
\caption{MCMC DIC Values for GNP Application Study} \label{gnp_dic_table}
\centering
\begin{tabular}[t]{l|l|r}
\hline
Model & Method & DIC\\
\hline
ARFIMA(0,d,0) & MCMC & 1008.25\\
\hline
ARFIMA(1,d,1) & MCMC & 1008.28\\
\hline
ARFIMA(0,d,1) & MCMC & 1009.95\\
\hline
ARFIMA(1,d,0) & MCMC & 1010.10\\
\hline
\end{tabular}
\end{table}

Overall, all model fits returned similar long-memory estimates that were consistent between the ABC and MCMC methods. The non-zero estimates of $d$ were in agreement with previous work on GNP datasets. However, model selection was inconclusive in deciding the order of the ARFIMA model, due to the very similar DIC values.

\section{Discussion and Concluding Remarks}
\label{section6}

In this paper, we presented a comparison between MCMC and ABC methods for estimating long-memory and ARMA parameters of ARFIMA models, up to orders $p,q \leq 1$. These methods were tested in an extensive simulation study with parameters of varying magnitude. All algorithms effectively determined the presence and approximate magnitude of long-memory and provided good estimates of $(d,\phi,\theta,\sigma^2)$. The novel MCMC Filter method, which we proposed as an alternative to the MCMC method by \cite{RN3}, filters out the ARMA and long-memory processes and yielded the best estimates. The ABC methods also performed well, with both ABC FP and ABC 20P providing similar results, effectively capturing the presence and strength of long-memory as well as the short memory ARMA components, making ABC a good approximation method to estimate ARFIMA parameters.

For target series with large $n$, ABC methods offer a computational advantage over MCMC. Although ABC requires significantly more simulations, making it slower for shorter series, the complex matrix operations in the MCMC methods make them much more computationally intensive for longer series. Additionally, the inherent parallelizability of ABC is particularly attractive compared to the sequential nature of MCMC, making ABC more efficient in distributed computing environments.

Another issue that emerged in this work regarding the MCMC methods was the high degree of serial correlation in the chains, particularly for the MCMC Filter. This correlation required significant thinning to achieve high effective sample sizes, which substantially increased the computational burden. ABC, on the other hand, did not face such issues of serial correlation due to the independence of the posterior draws.

The MCMC and ABC methods were also compared to the standard frequentist approach using MLE ARFIMA estimates. For frequentist estimates, the coverage was significantly worse than that of the Bayesian methods, particularly for $d$ and $\phi$. The frequentist estimates of $d$ and $\phi$ performed well for low-magnitude $\phi=0.2$ and ARFIMA$(0,d,1)$ models; however, they performed poorly in the presence of a large $\phi=0.5$.  

One of the most prominent issues that could be addressed in future work is improving the computational efficiency of both MCMC and ABC methods. This work demonstrates the usefulness of an ABC approach to modeling long-memory using a rejection sampling algorithm; however, several more computationally efficient algorithms can be employed in its place to expedite the sampling procedure significantly. Having demonstrated the viability of the ABC approach, we consider it an appealing alternative to the MCMC method, with significant potential for further development.

It is important to note that this work does not address the problem of model misspecification and assumes that an ARFIMA model with Gaussian innovations is a good representation of the data. If this assumption does not hold, alternative semiparametric or nonparametric approaches may be preferable; see, for example, \cite{grazian2024bayesian}.

\section{Supplementary Information}
\texttt{R} code for ABC and MCMC simulation studies is available at \\ \url{https://github.com/J-Cohen-Gabor/Bayesian-ARFIMA-Estimation}

\bibliography{sn-bibliography}

\newpage
\begin{appendices}

\onecolumn
\section{Simulation Study Results}
\label{app:simulations}

\subsection{Inferential results}
Columns with the smallest RMSE for each parameter are bold.

\subsubsection{ARFIMA$(1,d,1)$ Results} \label{appendix:arfima1d1}
\ 
\begin{table}[H]
\centering
\caption{ARFIMA(1,d,1) Simulation Results for $(d=0.05,\phi=0.2,\theta=0.2)$}
\centering
\begin{tabular}[t]{l|l|r|r|r|l|r|>{}r|l}
\hline
Par & Model & Mean & LB & UB & Coverage & SD & RMSE & ESS/$\bar{m}$\\
\hline
$d$ & MCMC & 0.03998 & -0.05180 & 0.12937 & 90\% & 0.05524 & \textcolor{black}{0.05792} & 723\\
\hline
$\phi$ & MCMC & 0.19684 & -0.01356 & 0.39201 & 88\% & 0.12385 & \textcolor{black}{0.12406} & 762\\
\hline
$\theta$ & MCMC & 0.21214 & 0.06901 & 0.35995 & 88\% & 0.08897 & \textcolor{black}{\textbf{0.08697}} & 880\\
\hline
$\sigma^2$ & MCMC & 1.00743 & 0.93773 & 1.08144 & 89\% & 0.04389 & \textcolor{black}{0.04537} & 959\\
\hline
$d$ & MCMC Filter & 0.04488 & -0.04358 & 0.13187 & 90\% & 0.05355 & \textcolor{black}{0.05589} & 591\\
\hline
$\phi$ & MCMC Filter & 0.18576 & -0.02118 & 0.37652 & 90\% & 0.12140 & \textcolor{black}{0.12320} & 648\\
\hline
$\theta$ & MCMC Filter & 0.21713 & 0.07419 & 0.36458 & 87\% & 0.08865 & \textcolor{black}{0.08770} & 839\\
\hline
\end{tabular}
\end{table}

\newpage
\begin{table}[H]
\centering
\caption{ARFIMA(1,d,1) Simulation Results for $(d=0.05,\phi=0.5,\theta=0.5)$}
\centering
\begin{tabular}[t]{l|l|r|r|r|l|r|>{}r|l}
\hline
Par & Model & Mean & LB & UB & Coverage & SD & RMSE & ESS/$\bar{m}$\\
\hline
$d$ & MCMC & 0.02523 & -0.10938 & 0.15190 & 90\% & 0.08081 & \textcolor{black}{0.08908} & 459\\
\hline
$\phi$ & MCMC & 0.51471 & 0.37164 & 0.65641 & 91\% & 0.08770 & \textcolor{black}{0.09089} & 484\\
\hline
$\theta$ & MCMC & 0.50812 & 0.44985 & 0.56395 & 88\% & 0.03480 & \textcolor{black}{\textbf{0.03431}} & 890\\
\hline
$\sigma^2$ & MCMC & 1.00732 & 0.93734 & 1.08170 & 90\% & 0.04397 & \textcolor{black}{0.04554} & 949\\
\hline
$d$ & MCMC Filter & 0.03420 & -0.08656 & 0.15154 & 88\% & 0.07333 & \textcolor{black}{0.09280} & 377\\
\hline
$\phi$ & MCMC Filter & 0.50389 & 0.37068 & 0.63309 & 88\% & 0.08064 & \textcolor{black}{0.09378} & 396\\
\hline
$\theta$ & MCMC Filter & 0.50858 & 0.45067 & 0.56407 & 88\% & 0.03462 & \textcolor{black}{0.03495} & 887\\
\hline
\end{tabular}
\end{table}

\newpage
\begin{table}[H]
\centering
\caption{ARFIMA(1,d,1) Simulation Results for $(d=0.1,\phi=0.2,\theta=0.2)$}
\centering
\begin{tabular}[t]{l|l|r|r|r|l|r|>{}r|l}
\hline
Par & Model & Mean & LB & UB & Coverage & SD & RMSE & ESS/$\bar{m}$\\
\hline
$d$ & MCMC & 0.08916 & -0.00144 & 0.17761 & 92\% & 0.05462 & \textcolor{black}{0.05779} & 719\\
\hline
$\phi$ & MCMC & 0.19858 & -0.01037 & 0.39196 & 88\% & 0.12286 & \textcolor{black}{0.12290} & 752\\
\hline
$\theta$ & MCMC & 0.21112 & 0.06891 & 0.35848 & 89\% & 0.08860 & \textcolor{black}{\textbf{0.08656}} & 865\\
\hline
$\sigma^2$ & MCMC & 1.00786 & 0.93786 & 1.08220 & 88\% & 0.04406 & \textcolor{black}{0.04568} & 959\\
\hline
$d$ & MCMC Filter & 0.09348 & 0.00567 & 0.17986 & 90\% & 0.05330 & \textcolor{black}{0.05499} & 592\\
\hline
$\phi$ & MCMC Filter & 0.18869 & -0.01651 & 0.37798 & 88\% & 0.12059 & \textcolor{black}{\textbf{0.12095}} & 658\\
\hline
$\theta$ & MCMC Filter & 0.21549 & 0.07310 & 0.36212 & 88\% & 0.08812 & \textcolor{black}{0.08695} & 854\\
\hline
\end{tabular}
\end{table}

\newpage
\begin{table}[H]
\centering
\caption{ARFIMA(1,d,1) Simulation Results for $(d=0.1,\phi=0.5,\theta=0.5)$}
\centering
\begin{tabular}[t]{l|l|r|r|r|l|r|>{}r|l}
\hline
Par & Model & Mean & LB & UB & Coverage & SD & RMSE & ESS/$\bar{m}$\\
\hline
$d$ & MCMC & 0.07282 & -0.06087 & 0.20032 & 90\% & 0.08055 & \textcolor{black}{0.09272} & 475\\
\hline
$\phi$ & MCMC & 0.51678 & 0.37301 & 0.65762 & 89\% & 0.08731 & \textcolor{black}{\textbf{0.09302}} & 493\\
\hline
$\theta$ & MCMC & 0.50828 & 0.45006 & 0.56412 & 88\% & 0.03485 & \textcolor{black}{\textbf{0.03462}} & 868\\
\hline
$\sigma^2$ & MCMC & 1.00727 & 0.93716 & 1.08156 & 89\% & 0.04400 & \textcolor{black}{0.04536} & 966\\
\hline
$d$ & MCMC Filter & 0.08250 & -0.03731 & 0.20031 & 88\% & 0.07298 & \textcolor{black}{0.09386} & 382\\
\hline
$\phi$ & MCMC Filter & 0.50567 & 0.37229 & 0.63340 & 88\% & 0.08017 & \textcolor{black}{0.09423} & 399\\
\hline
$\theta$ & MCMC Filter & 0.50854 & 0.45056 & 0.56409 & 88\% & 0.03460 & \textcolor{black}{0.03483} & 892\\
\hline
\end{tabular}
\end{table}

\newpage
\begin{table}[H]
\centering
\caption{ARFIMA(1,d,1) Simulation Results for $(d=0.2,\phi=0.2,\theta=0.2)$}
\centering
\begin{tabular}[t]{l|l|r|r|r|l|r|>{}r|l}
\hline
Par & Model & Mean & LB & UB & Coverage & SD & RMSE & ESS/$\bar{m}$\\
\hline
$d$ & MCMC & 0.18730 & 0.09483 & 0.27496 & 92\% & 0.05492 & \textcolor{black}{0.05738} & 732\\
\hline
$\phi$ & MCMC & 0.20140 & -0.00637 & 0.39535 & 87\% & 0.12260 & \textcolor{black}{0.12187} & 763\\
\hline
$\theta$ & MCMC & 0.21000 & 0.06802 & 0.35731 & 88\% & 0.08824 & \textcolor{black}{\textbf{0.08605}} & 864\\
\hline
$\sigma^2$ & MCMC & 1.00750 & 0.93776 & 1.08180 & 89\% & 0.04394 & \textcolor{black}{0.04565} & 956\\
\hline
$d$ & MCMC Filter & 0.18987 & 0.10156 & 0.27610 & 90\% & 0.05331 & \textcolor{black}{\textbf{0.05634}} & 599\\
\hline
$\phi$ & MCMC Filter & 0.19468 & -0.00946 & 0.38409 & 88\% & 0.12021 & \textcolor{black}{\textbf{0.12045}} & 657\\
\hline
$\theta$ & MCMC Filter & 0.21319 & 0.07202 & 0.35883 & 88\% & 0.08763 & \textcolor{black}{0.08610} & 845\\
\hline
\end{tabular}
\end{table}

\newpage
\begin{table}[H]
\centering
\caption{ARFIMA(1,d,1) Simulation Results for $(d=0.2,\phi=0.5,\theta=0.5)$}
\centering
\begin{tabular}[t]{l|l|r|r|r|l|r|>{}r|l}
\hline
Par & Model & Mean & LB & UB & Coverage & SD & RMSE & ESS/$\bar{m}$\\
\hline
$d$ & MCMC & 0.16983 & 0.03454 & 0.29313 & 90\% & 0.08006 & \textcolor{black}{0.09120} & 479\\
\hline
$\phi$ & MCMC & 0.51983 & 0.38093 & 0.66136 & 90\% & 0.08645 & \textcolor{black}{0.09126} & 499\\
\hline
$\theta$ & MCMC & 0.50831 & 0.44968 & 0.56437 & 90\% & 0.03501 & \textcolor{black}{0.03439} & 865\\
\hline
$\sigma^2$ & MCMC & 1.00720 & 0.93704 & 1.08169 & 88\% & 0.04409 & \textcolor{black}{0.04546} & 958\\
\hline
$d$ & MCMC Filter & 0.17722 & 0.05849 & 0.29443 & 90\% & 0.07190 & \textcolor{black}{0.08970} & 390\\
\hline
$\phi$ & MCMC Filter & 0.51167 & 0.37928 & 0.63833 & 88\% & 0.07899 & \textcolor{black}{\textbf{0.09093}} & 405\\
\hline
$\theta$ & MCMC Filter & 0.50813 & 0.45042 & 0.56372 & 88\% & 0.03450 & \textcolor{black}{\textbf{0.03407}} & 892\\
\hline
\end{tabular}
\end{table}

\newpage
\begin{table}[H]
\centering
\caption{ARFIMA(1,d,1) Simulation Results for $(d=0.3,\phi=0.2,\theta=0.2)$}
\centering
\begin{tabular}[t]{l|l|r|r|r|l|r|>{}r|l}
\hline
Par & Model & Mean & LB & UB & Coverage & SD & RMSE & ESS/$\bar{m}$\\
\hline
$d$ & MCMC & 0.28197 & 0.18947 & 0.36894 & 90\% & 0.05501 & \textcolor{black}{0.05943} & 740\\
\hline
$\phi$ & MCMC & 0.20995 & 0.00587 & 0.40239 & 90\% & 0.12140 & \textcolor{black}{0.12131} & 769\\
\hline
$\theta$ & MCMC & 0.20651 & 0.06559 & 0.35151 & 89\% & 0.08748 & \textcolor{black}{0.08461} & 869\\
\hline
$\sigma^2$ & MCMC & 1.00737 & 0.93748 & 1.08194 & 90\% & 0.04399 & \textcolor{black}{0.04559} & 963\\
\hline
$d$ & MCMC Filter & 0.28396 & 0.19864 & 0.36705 & 92\% & 0.05143 & \textcolor{black}{\textbf{0.05336}} & 603\\
\hline
$\phi$ & MCMC Filter & 0.20636 & 0.00732 & 0.39064 & 88\% & 0.11717 & \textcolor{black}{\textbf{0.11545}} & 669\\
\hline
$\theta$ & MCMC Filter & 0.20756 & 0.06795 & 0.35093 & 89\% & 0.08622 & \textcolor{black}{\textbf{0.08414}} & 847\\
\hline
\end{tabular}
\end{table}

\newpage
\begin{table}[H]
\centering
\caption{ARFIMA(1,d,1) Simulation Results for $(d=0.3,\phi=0.5,\theta=0.5)$}
\centering
\begin{tabular}[t]{l|l|r|r|r|l|r|>{}r|l}
\hline
Par & Model & Mean & LB & UB & Coverage & SD & RMSE & ESS/$\bar{m}$\\
\hline
$d$ & MCMC & 0.26162 & 0.13006 & 0.38263 & 90\% & 0.07760 & \textcolor{black}{0.09315} & 495\\
\hline
$\phi$ & MCMC & 0.52812 & 0.39169 & 0.66536 & 90\% & 0.08402 & \textcolor{black}{0.09135} & 514\\
\hline
$\theta$ & MCMC & 0.50804 & 0.44973 & 0.56376 & 89\% & 0.03497 & \textcolor{black}{0.03400} & 876\\
\hline
$\sigma^2$ & MCMC & 1.00709 & 0.93736 & 1.08128 & 90\% & 0.04395 & \textcolor{black}{0.04546} & 955\\
\hline
$d$ & MCMC Filter & 0.26440 & 0.14986 & 0.37413 & 88\% & 0.06835 & \textcolor{black}{\textbf{0.09193}} & 420\\
\hline
$\phi$ & MCMC Filter & 0.52543 & 0.40045 & 0.64756 & 86\% & 0.07533 & \textcolor{black}{0.09061} & 435\\
\hline
$\theta$ & MCMC Filter & 0.50763 & 0.45016 & 0.56251 & 90\% & 0.03422 & \textcolor{black}{\textbf{0.03365}} & 891\\
\hline
\end{tabular}
\end{table}

\newpage
\begin{table}[H]
\centering
\caption{ARFIMA(1,d,1) Simulation Results for $(d=0.4,\phi=0.2,\theta=0.2)$}
\centering
\begin{tabular}[t]{l|l|r|r|r|l|r|>{}r|l}
\hline
Par & Model & Mean & LB & UB & Coverage & SD & RMSE & ESS/$\bar{m}$\\
\hline
$d$ & MCMC & 0.36628 & 0.28305 & 0.44280 & 91\% & 0.04933 & \textcolor{black}{0.08094} & 777\\
\hline
$\phi$ & MCMC & 0.23552 & 0.04668 & 0.41465 & 85\% & 0.11294 & \textcolor{black}{0.12481} & 798\\
\hline
$\theta$ & MCMC & 0.19645 & 0.06050 & 0.33467 & 90\% & 0.08364 & \textcolor{black}{0.08208} & 893\\
\hline
$\sigma^2$ & MCMC & 1.00694 & 0.93700 & 1.08073 & 90\% & 0.04393 & \textcolor{black}{0.04516} & 952\\
\hline
$d$ & MCMC Filter & 0.36458 & 0.28560 & 0.43446 & 91\% & 0.04567 & \textcolor{black}{\textbf{0.05549}} & 672\\
\hline
$\phi$ & MCMC Filter & 0.24216 & 0.06332 & 0.41281 & 86\% & 0.10671 & \textcolor{black}{\textbf{0.11205}} & 732\\
\hline
$\theta$ & MCMC Filter & 0.19149 & 0.05882 & 0.32628 & 88\% & 0.08154 & \textcolor{black}{\textbf{0.08098}} & 899\\
\hline
\end{tabular}
\end{table}

\newpage
\begin{table}[H]
\centering
\caption{ARFIMA(1,d,1) Simulation Results for $(d=0.4,\phi=0.5,\theta=0.5)$}
\centering
\begin{tabular}[t]{l|l|r|r|r|l|r|>{}r|l}
\hline
Par & Model & Mean & LB & UB & Coverage & SD & RMSE & ESS/$\bar{m}$\\
\hline
$d$ & MCMC & 0.34013 & 0.21896 & 0.44492 & 88\% & 0.07062 & \textcolor{black}{0.09890} & 556\\
\hline
$\phi$ & MCMC & 0.55009 & 0.42884 & 0.67629 & 87\% & 0.07682 & \textcolor{black}{\textbf{0.09385}} & 575\\
\hline
$\theta$ & MCMC & 0.50772 & 0.45027 & 0.56288 & 90\% & 0.03451 & \textcolor{black}{0.03376} & 873\\
\hline
$\sigma^2$ & MCMC & 1.00682 & 0.93742 & 1.08084 & 89\% & 0.04388 & \textcolor{black}{0.04517} & 950\\
\hline
$d$ & MCMC Filter & 0.33515 & 0.23076 & 0.42734 & 84\% & 0.06036 & \textcolor{black}{0.10031} & 482\\
\hline
$\phi$ & MCMC Filter & 0.55604 & 0.44743 & 0.66757 & 82\% & 0.06719 & \textcolor{black}{0.09567} & 499\\
\hline
$\theta$ & MCMC Filter & 0.50712 & 0.45041 & 0.56100 & 90\% & 0.03373 & \textcolor{black}{\textbf{0.03318}} & 887\\
\hline
\end{tabular}
\end{table}

\newpage
\subsubsection{ARFIMA$(1,d,0)$ Results} \label{appendix:arfima1d0}
\ 

\begin{table}[H]
\centering
\caption{ARFIMA(1,d,0) Simulation Results for $(d=0.05,\phi=0.2)$}
\centering
\begin{tabular}[t]{l|l|r|r|r|l|r|>{}r|l}
\hline
Par & Model & Mean & LB & UB & Coverage & SD & RMSE & ESS/$\bar{m}$\\
\hline
$d$ & MCMC & 0.03829 & -0.04287 & 0.11868 & 92\% & 0.04934 & \textcolor{black}{0.05057} & 949\\
\hline
$\phi$ & MCMC & 0.21334 & 0.11457 & 0.31679 & 92\% & 0.06159 & \textcolor{black}{0.05930} & 949\\
\hline
$\sigma^2$ & MCMC & 1.00787 & 0.93774 & 1.08242 & 89\% & 0.04410 & \textcolor{black}{0.04568} & 953\\
\hline
$d$ & MCMC Filter & 0.04168 & -0.03784 & 0.12110 & 90\% & 0.04853 & \textcolor{black}{0.04907} & 706\\
\hline
$\phi$ & MCMC Filter & 0.20830 & 0.11193 & 0.30831 & 92\% & 0.05994 & \textcolor{black}{0.05740} & 707\\
\hline
$\sigma^2$ & MCMC Filter & 1.00796 & 0.93786 & 1.08218 & 90\% & 0.04402 & \textcolor{black}{0.04579} & 961\\
\hline
$d$ & ABC LogP-Q0.005 & 0.03294 & -0.13484 & 0.19981 & 100\% & 0.10618 & \textcolor{black}{0.05440} & 1128\\
\hline
\end{tabular}
\end{table}

\newpage
\begin{table}[H]
\centering
\caption{ARFIMA(1,d,0) Simulation Results for $(d=0.05,\phi=0.5)$}
\centering
\begin{tabular}[t]{l|l|r|r|r|l|r|>{}r|l}
\hline
Par & Model & Mean & LB & UB & Coverage & SD & RMSE & ESS/$\bar{m}$\\
\hline
$d$ & MCMC & 0.02735 & -0.10017 & 0.14854 & 91\% & 0.07629 & \textcolor{black}{0.07835} & 806\\
\hline
$\phi$ & MCMC & 0.51847 & 0.38788 & 0.65302 & 92\% & 0.08115 & \textcolor{black}{0.08016} & 816\\
\hline
$\sigma^2$ & MCMC & 1.00763 & 0.93793 & 1.08211 & 89\% & 0.04394 & \textcolor{black}{0.04573} & 959\\
\hline
$d$ & MCMC Filter & 0.03657 & -0.08289 & 0.15395 & 90\% & 0.07198 & \textcolor{black}{0.08148} & 425\\
\hline
$\phi$ & MCMC Filter & 0.50728 & 0.38206 & 0.63204 & 90\% & 0.07607 & \textcolor{black}{0.08266} & 429\\
\hline
$\sigma^2$ & MCMC Filter & 1.00771 & 0.93792 & 1.08215 & 90\% & 0.04395 & \textcolor{black}{0.04587} & 952\\
\hline
$d$ & ABC LogP-Q0.005 & 0.01565 & -0.18588 & 0.19888 & 97\% & 0.12080 & \textcolor{black}{0.07975} & 1077\\
\hline
\end{tabular}
\end{table}

\newpage
\begin{table}[H]
\centering
\caption{ARFIMA(1,d,0) Simulation Results for $(d=0.1,\phi=0.2)$}
\centering
\begin{tabular}[t]{l|l|r|r|r|l|r|>{}r|l}
\hline
Par & Model & Mean & LB & UB & Coverage & SD & RMSE & ESS/$\bar{m}$\\
\hline
$d$ & MCMC & 0.08794 & 0.00697 & 0.16840 & 92\% & 0.04932 & \textcolor{black}{0.05031} & 956\\
\hline
$\phi$ & MCMC & 0.21360 & 0.11516 & 0.31668 & 93\% & 0.06153 & \textcolor{black}{0.05888} & 948\\
\hline
$\sigma^2$ & MCMC & 1.00804 & 0.93832 & 1.08234 & 89\% & 0.04397 & \textcolor{black}{0.04575} & 950\\
\hline
$d$ & MCMC Filter & 0.09097 & 0.01173 & 0.17019 & 92\% & 0.04825 & \textcolor{black}{0.04866} & 724\\
\hline
$\phi$ & MCMC Filter & 0.20901 & 0.11270 & 0.30906 & 90\% & 0.05981 & \textcolor{black}{0.05737} & 724\\
\hline
$\sigma^2$ & MCMC Filter & 1.00788 & 0.93818 & 1.08199 & 88\% & 0.04393 & \textcolor{black}{0.04578} & 965\\
\hline
$d$ & ABC LogP-Q0.005 & 0.07833 & -0.09487 & 0.24814 & 100\% & 0.11619 & \textcolor{black}{0.05471} & 1111\\
\hline
\end{tabular}
\end{table}

\newpage
\begin{table}[H]
\centering
\caption{ARFIMA(1,d,0) Simulation Results for $(d=0.1,\phi=0.5)$}
\centering
\begin{tabular}[t]{l|l|r|r|r|l|r|>{}r|l}
\hline
Par & Model & Mean & LB & UB & Coverage & SD & RMSE & ESS/$\bar{m}$\\
\hline
$d$ & MCMC & 0.07689 & -0.04963 & 0.19715 & 92\% & 0.07533 & \textcolor{black}{0.07809} & 810\\
\hline
$\phi$ & MCMC & 0.51889 & 0.38930 & 0.65230 & 91\% & 0.08020 & \textcolor{black}{0.07996} & 817\\
\hline
$\sigma^2$ & MCMC & 1.00756 & 0.93776 & 1.08154 & 89\% & 0.04387 & \textcolor{black}{0.04587} & 952\\
\hline
$d$ & MCMC Filter & 0.08636 & -0.03182 & 0.20254 & 90\% & 0.07177 & \textcolor{black}{0.07727} & 429\\
\hline
$\phi$ & MCMC Filter & 0.50770 & 0.38306 & 0.63142 & 90\% & 0.07592 & \textcolor{black}{0.07925} & 431\\
\hline
$\sigma^2$ & MCMC Filter & 1.00761 & 0.93803 & 1.08167 & 88\% & 0.04390 & \textcolor{black}{0.04579} & 955\\
\hline
$d$ & ABC LogP-Q0.005 & 0.06122 & -0.13925 & 0.24150 & 97\% & 0.11967 & \textcolor{black}{0.08053} & 1037\\
\hline
\end{tabular}
\end{table}

\newpage
\begin{table}[H]
\centering
\caption{ARFIMA(1,d,0) Simulation Results for $(d=0.2,\phi=0.2)$}
\centering
\begin{tabular}[t]{l|l|r|r|r|l|r|>{}r|l}
\hline
Par & Model & Mean & LB & UB & Coverage & SD & RMSE & ESS/$\bar{m}$\\
\hline
$d$ & MCMC & 0.18605 & 0.10406 & 0.26622 & 92\% & 0.04970 & \textcolor{black}{0.05189} & 940\\
\hline
$\phi$ & MCMC & 0.21547 & 0.11727 & 0.31955 & 94\% & 0.06207 & \textcolor{black}{0.06052} & 941\\
\hline
$\sigma^2$ & MCMC & 1.00773 & 0.93790 & 1.08172 & 88\% & 0.04390 & \textcolor{black}{0.04580} & 958\\
\hline
$d$ & MCMC Filter & 0.18842 & 0.10866 & 0.26680 & 92\% & 0.04824 & \textcolor{black}{0.04917} & 728\\
\hline
$\phi$ & MCMC Filter & 0.21164 & 0.11621 & 0.31257 & 92\% & 0.05999 & \textcolor{black}{0.05788} & 717\\
\hline
$\sigma^2$ & MCMC Filter & 1.00759 & 0.93731 & 1.08166 & 89\% & 0.04405 & \textcolor{black}{0.04562} & 964\\
\hline
$d$ & ABC LogP-Q0.005 & 0.15173 & -0.17724 & 0.33567 & 100\% & 0.16043 & \textcolor{black}{0.06557} & 1090\\
\hline
\end{tabular}
\end{table}

\newpage
\begin{table}[H]
\centering
\caption{ARFIMA(1,d,0) Simulation Results for $(d=0.2,\phi=0.5)$}
\centering
\begin{tabular}[t]{l|l|r|r|r|l|r|>{}r|l}
\hline
Par & Model & Mean & LB & UB & Coverage & SD & RMSE & ESS/$\bar{m}$\\
\hline
$d$ & MCMC & 0.17413 & 0.05034 & 0.29264 & 90\% & 0.07417 & \textcolor{black}{0.07840} & 815\\
\hline
$\phi$ & MCMC & 0.52153 & 0.39329 & 0.65252 & 90\% & 0.07917 & \textcolor{black}{0.08016} & 817\\
\hline
$\sigma^2$ & MCMC & 1.00742 & 0.93755 & 1.08207 & 89\% & 0.04409 & \textcolor{black}{0.04577} & 957\\
\hline
$d$ & MCMC Filter & 0.18064 & 0.06614 & 0.29385 & 89\% & 0.06971 & \textcolor{black}{0.07729} & 441\\
\hline
$\phi$ & MCMC Filter & 0.51381 & 0.39149 & 0.63443 & 90\% & 0.07415 & \textcolor{black}{0.07918} & 438\\
\hline
$\sigma^2$ & MCMC Filter & 1.00752 & 0.93806 & 1.08137 & 88\% & 0.04385 & \textcolor{black}{0.04575} & 955\\
\hline
$d$ & ABC LogP-Q0.005 & 0.15198 & -0.05004 & 0.32841 & 96\% & 0.11943 & \textcolor{black}{0.08381} & 992\\
\hline
\end{tabular}
\end{table}

\newpage
\begin{table}[H]
\centering
\caption{ARFIMA(1,d,0) Simulation Results for $(d=0.3,\phi=0.2)$}
\centering
\begin{tabular}[t]{l|l|r|r|r|l|r|>{}r|l}
\hline
Par & Model & Mean & LB & UB & Coverage & SD & RMSE & ESS/$\bar{m}$\\
\hline
$d$ & MCMC & 0.28235 & 0.19914 & 0.36135 & 90\% & 0.04955 & \textcolor{black}{0.05105} & 939\\
\hline
$\phi$ & MCMC & 0.21891 & 0.12118 & 0.32424 & 93\% & 0.06199 & \textcolor{black}{0.06067} & 934\\
\hline
$\sigma^2$ & MCMC & 1.00760 & 0.93795 & 1.08199 & 90\% & 0.04403 & \textcolor{black}{0.04563} & 958\\
\hline
$d$ & MCMC Filter & 0.28320 & 0.20452 & 0.36059 & 91\% & 0.04766 & \textcolor{black}{0.04964} & 739\\
\hline
$\phi$ & MCMC Filter & 0.21723 & 0.12167 & 0.31732 & 92\% & 0.05970 & \textcolor{black}{0.05866} & 730\\
\hline
$\sigma^2$ & MCMC Filter & 1.00741 & 0.93775 & 1.08141 & 90\% & 0.04392 & \textcolor{black}{0.04564} & 953\\
\hline
$d$ & ABC LogP-Q0.005 & 0.20149 & -0.37813 & 0.40695 & 100\% & 0.21735 & \textcolor{black}{0.10532} & 1045\\
\hline
\end{tabular}
\end{table}

\newpage
\begin{table}[H]
\centering
\caption{ARFIMA(1,d,0) Simulation Results for $(d=0.3,\phi=0.5)$}
\centering
\begin{tabular}[t]{l|l|r|r|r|l|r|>{}r|l}
\hline
Par & Model & Mean & LB & UB & Coverage & SD & RMSE & ESS/$\bar{m}$\\
\hline
$d$ & MCMC & 0.26634 & 0.14642 & 0.38056 & 90\% & 0.07183 & \textcolor{black}{\textbf{0.07861}} & 834\\
\hline
$\phi$ & MCMC & 0.52924 & 0.40504 & 0.65607 & 90\% & 0.07694 & \textcolor{black}{0.08010} & 833\\
\hline
$\sigma^2$ & MCMC & 1.00723 & 0.93719 & 1.08132 & 88\% & 0.04397 & \textcolor{black}{0.04555} & 959\\
\hline
$d$ & MCMC Filter & 0.26790 & 0.15771 & 0.37483 & 90\% & 0.06657 & \textcolor{black}{0.08013} & 473\\
\hline
$\phi$ & MCMC Filter & 0.52693 & 0.41108 & 0.64356 & 89\% & 0.07125 & \textcolor{black}{0.08070} & 472\\
\hline
$\sigma^2$ & MCMC Filter & 1.00715 & 0.93722 & 1.08123 & 89\% & 0.04397 & \textcolor{black}{0.04551} & 955\\
\hline
$d$ & ABC LogP-Q0.005 & 0.24146 & 0.03353 & 0.41147 & 96\% & 0.11828 & \textcolor{black}{0.08729} & 1041\\
\hline
\end{tabular}
\end{table}

\newpage
\begin{table}[H]
\centering
\caption{ARFIMA(1,d,0) Simulation Results for $(d=0.4,\phi=0.2)$}
\centering
\begin{tabular}[t]{l|l|r|r|r|l|r|>{}r|l}
\hline
Par & Model & Mean & LB & UB & Coverage & SD & RMSE & ESS/$\bar{m}$\\
\hline
$d$ & MCMC & 0.37228 & 0.29342 & 0.44379 & 88\% & 0.04565 & \textcolor{black}{0.05102} & 942\\
\hline
$\phi$ & MCMC & 0.22932 & 0.13711 & 0.33071 & 90\% & 0.05890 & \textcolor{black}{0.06111} & 940\\
\hline
$\sigma^2$ & MCMC & 1.00715 & 0.93750 & 1.08137 & 88\% & 0.04397 & \textcolor{black}{0.04561} & 957\\
\hline
$d$ & MCMC Filter & 0.36818 & 0.29692 & 0.43327 & 89\% & 0.04202 & \textcolor{black}{0.04955} & 770\\
\hline
$\phi$ & MCMC Filter & 0.23349 & 0.14615 & 0.32772 & 90\% & 0.05562 & \textcolor{black}{0.05969} & 769\\
\hline
$\sigma^2$ & MCMC Filter & 1.00733 & 0.93726 & 1.08177 & 89\% & 0.04407 & \textcolor{black}{0.04528} & 952\\
\hline
$d$ & ABC LogP-Q0.005 & 0.23306 & -0.36813 & 0.46355 & 99\% & 0.26649 & \textcolor{black}{0.17045} & 1051\\
\hline
\end{tabular}
\end{table}

\newpage
\begin{table}[H]
\centering
\caption{ARFIMA(1,d,0) Simulation Results for $(d=0.4,\phi=0.5)$}
\centering
\begin{tabular}[t]{l|l|r|r|r|l|r|>{}r|l}
\hline
Par & Model & Mean & LB & UB & Coverage & SD & RMSE & ESS/$\bar{m}$\\
\hline
$d$ & MCMC & 0.34562 & 0.23295 & 0.44641 & 90\% & 0.06547 & \textcolor{black}{0.08466} & 858\\
\hline
$\phi$ & MCMC & 0.55003 & 0.43806 & 0.66975 & 88\% & 0.07097 & \textcolor{black}{0.08478} & 856\\
\hline
$\sigma^2$ & MCMC & 1.00692 & 0.93711 & 1.08106 & 89\% & 0.04394 & \textcolor{black}{0.04541} & 954\\
\hline
$d$ & MCMC Filter & 0.33929 & 0.23697 & 0.42905 & 86\% & 0.05907 & \textcolor{black}{0.08968} & 529\\
\hline
$\phi$ & MCMC Filter & 0.55632 & 0.45505 & 0.66446 & 85\% & 0.06426 & \textcolor{black}{0.08876} & 533\\
\hline
$\sigma^2$ & MCMC Filter & 1.00701 & 0.93718 & 1.08096 & 88\% & 0.04397 & \textcolor{black}{0.04532} & 956\\
\hline
$d$ & ABC LogP-Q0.005 & 0.30598 & 0.05376 & 0.46444 & 96\% & 0.12842 & \textcolor{black}{0.10435} & 1144\\
\hline
\end{tabular}
\end{table}

\newpage
\subsubsection{ARFIMA$(0,d,1)$ Results} \label{appendix:arfima0d1}
\ 
\begin{table}[H]
\centering
\caption{ARFIMA(0,d,1) Simulation Results for $(d=0.05,\theta=0.2)$}
\centering
\begin{tabular}[t]{l|l|r|r|r|l|r|>{}r|l}
\hline
Par & Model & Mean & LB & UB & Coverage & SD & RMSE & ESS/$\bar{m}$\\
\hline
$d$ & MCMC & 0.04853 & -0.00699 & 0.10773 & 92\% & 0.03506 & \textcolor{black}{0.03386} & 955\\
\hline
$\theta$ & MCMC & 0.20178 & 0.12868 & 0.27124 & 92\% & 0.04358 & \textcolor{black}{0.04045} & 961\\
\hline
$\sigma^2$ & MCMC & 1.00790 & 0.93789 & 1.08187 & 90\% & 0.04406 & \textcolor{black}{0.04567} & 954\\
\hline
$d$ & MCMC Filter & 0.04918 & -0.00621 & 0.10823 & 92\% & 0.03493 & \textcolor{black}{0.03395} & 930\\
\hline
$\theta$ & MCMC Filter & 0.20029 & 0.12742 & 0.26980 & 91\% & 0.04348 & \textcolor{black}{0.04090} & 928\\
\hline
$\sigma^2$ & MCMC Filter & 1.00796 & 0.93802 & 1.08243 & 90\% & 0.04400 & \textcolor{black}{0.04556} & 962\\
\hline
$d$ & ABC LogP-Q0.005 & 0.04525 & -0.11411 & 0.20386 & 100\% & 0.09671 & \textcolor{black}{0.03690} & 1842\\
\hline
\end{tabular}
\end{table}

\newpage
\begin{table}[H]
\centering
\caption{ARFIMA(0,d,1) Simulation Results for $(d=0.05,\theta=0.5)$}
\centering
\begin{tabular}[t]{l|l|r|r|r|l|r|>{}r|l}
\hline
Par & Model & Mean & LB & UB & Coverage & SD & RMSE & ESS/$\bar{m}$\\
\hline
$d$ & MCMC & 0.04869 & 0.00064 & 0.09890 & 92\% & 0.02992 & \textcolor{black}{0.02960} & 965\\
\hline
$\theta$ & MCMC & 0.50118 & 0.44559 & 0.55368 & 92\% & 0.03295 & \textcolor{black}{\textbf{0.03100}} & 958\\
\hline
$\sigma^2$ & MCMC & 1.00761 & 0.93772 & 1.08189 & 90\% & 0.04403 & \textcolor{black}{0.04549} & 951\\
\hline
$d$ & MCMC Filter & 0.04878 & 0.00094 & 0.09895 & 93\% & 0.02986 & \textcolor{black}{0.02957} & 957\\
\hline
$\theta$ & MCMC Filter & 0.50051 & 0.44511 & 0.55279 & 91\% & 0.03281 & \textcolor{black}{0.03143} & 960\\
\hline
$\sigma^2$ & MCMC Filter & 1.00760 & 0.93773 & 1.08179 & 88\% & 0.04395 & \textcolor{black}{0.04548} & 955\\
\hline
$d$ & ABC LogP-Q0.005 & 0.03943 & -0.11501 & 0.19302 & 100\% & 0.09393 & \textcolor{black}{0.03229} & 2247\\
\hline
\end{tabular}
\end{table}

\newpage
\begin{table}[H]
\centering
\caption{ARFIMA(0,d,1) Simulation Results for $(d=0.1,\theta=0.2)$}
\centering
\begin{tabular}[t]{l|l|r|r|r|l|r|>{}r|l}
\hline
Par & Model & Mean & LB & UB & Coverage & SD & RMSE & ESS/$\bar{m}$\\
\hline
$d$ & MCMC & 0.09823 & 0.04303 & 0.15753 & 93\% & 0.03488 & \textcolor{black}{0.03384} & 953\\
\hline
$\theta$ & MCMC & 0.20207 & 0.12903 & 0.27173 & 93\% & 0.04353 & \textcolor{black}{0.04062} & 955\\
\hline
$\sigma^2$ & MCMC & 1.00807 & 0.93794 & 1.08255 & 89\% & 0.04414 & \textcolor{black}{0.04557} & 954\\
\hline
$d$ & MCMC Filter & 0.09894 & 0.04399 & 0.15764 & 92\% & 0.03473 & \textcolor{black}{\textbf{0.03373}} & 917\\
\hline
$\theta$ & MCMC Filter & 0.20068 & 0.12828 & 0.27009 & 91\% & 0.04326 & \textcolor{black}{0.04078} & 926\\
\hline
$\sigma^2$ & MCMC Filter & 1.00779 & 0.93773 & 1.08214 & 89\% & 0.04402 & \textcolor{black}{0.04562} & 962\\
\hline
$d$ & ABC LogP-Q0.005 & 0.09401 & -0.06392 & 0.25098 & 100\% & 0.09580 & \textcolor{black}{0.03640} & 1832\\
\hline
\end{tabular}
\end{table}

\newpage
\begin{table}[H]
\centering
\caption{ARFIMA(0,d,1) Simulation Results for $(d=0.1,\theta=0.5)$}
\centering
\begin{tabular}[t]{l|l|r|r|r|l|r|>{}r|l}
\hline
Par & Model & Mean & LB & UB & Coverage & SD & RMSE & ESS/$\bar{m}$\\
\hline
$d$ & MCMC & 0.09856 & 0.05083 & 0.14874 & 92\% & 0.02988 & \textcolor{black}{0.02950} & 960\\
\hline
$\theta$ & MCMC & 0.50119 & 0.44598 & 0.55358 & 93\% & 0.03285 & \textcolor{black}{0.03113} & 953\\
\hline
$\sigma^2$ & MCMC & 1.00761 & 0.93754 & 1.08208 & 89\% & 0.04400 & \textcolor{black}{0.04542} & 955\\
\hline
$d$ & MCMC Filter & 0.09869 & 0.05095 & 0.14873 & 93\% & 0.02982 & \textcolor{black}{\textbf{0.02940}} & 956\\
\hline
$\theta$ & MCMC Filter & 0.50061 & 0.44526 & 0.55280 & 92\% & 0.03280 & \textcolor{black}{0.03143} & 952\\
\hline
$\sigma^2$ & MCMC Filter & 1.00777 & 0.93807 & 1.08220 & 89\% & 0.04398 & \textcolor{black}{0.04531} & 964\\
\hline
$d$ & ABC LogP-Q0.005 & 0.08513 & -0.06700 & 0.23372 & 100\% & 0.09177 & \textcolor{black}{0.03339} & 2163\\
\hline
\end{tabular}
\end{table}

\newpage
\begin{table}[H]
\centering
\caption{ARFIMA(0,d,1) Simulation Results for $(d=0.2,\theta=0.2)$}
\centering
\begin{tabular}[t]{l|l|r|r|r|l|r|>{}r|l}
\hline
Par & Model & Mean & LB & UB & Coverage & SD & RMSE & ESS/$\bar{m}$\\
\hline
$d$ & MCMC & 0.19728 & 0.14266 & 0.25587 & 94\% & 0.03454 & \textcolor{black}{0.03345} & 959\\
\hline
$\theta$ & MCMC & 0.20273 & 0.12997 & 0.27204 & 92\% & 0.04332 & \textcolor{black}{0.04033} & 957\\
\hline
$\sigma^2$ & MCMC & 1.00775 & 0.93763 & 1.08206 & 89\% & 0.04403 & \textcolor{black}{0.04558} & 953\\
\hline
$d$ & MCMC Filter & 0.19749 & 0.14275 & 0.25579 & 93\% & 0.03453 & \textcolor{black}{\textbf{0.03344}} & 925\\
\hline
$\theta$ & MCMC Filter & 0.20206 & 0.12974 & 0.27115 & 92\% & 0.04317 & \textcolor{black}{0.04045} & 916\\
\hline
$\sigma^2$ & MCMC Filter & 1.00775 & 0.93793 & 1.08182 & 90\% & 0.04393 & \textcolor{black}{0.04559} & 955\\
\hline
$d$ & ABC LogP-Q0.005 & 0.18574 & 0.03448 & 0.33285 & 100\% & 0.09095 & \textcolor{black}{0.03608} & 1750\\
\hline
\end{tabular}
\end{table}

\newpage
\begin{table}[H]
\centering
\caption{ARFIMA(0,d,1) Simulation Results for $(d=0.2,\theta=0.5)$}
\centering
\begin{tabular}[t]{l|l|r|r|r|l|r|>{}r|l}
\hline
Par & Model & Mean & LB & UB & Coverage & SD & RMSE & ESS/$\bar{m}$\\
\hline
$d$ & MCMC & 0.19775 & 0.15035 & 0.24762 & 92\% & 0.02965 & \textcolor{black}{\textbf{0.02931}} & 955\\
\hline
$\theta$ & MCMC & 0.50155 & 0.44628 & 0.55367 & 92\% & 0.03276 & \textcolor{black}{\textbf{0.03098}} & 953\\
\hline
$\sigma^2$ & MCMC & 1.00736 & 0.93770 & 1.08157 & 90\% & 0.04387 & \textcolor{black}{0.04535} & 957\\
\hline
$d$ & MCMC Filter & 0.19768 & 0.15042 & 0.24720 & 92\% & 0.02954 & \textcolor{black}{0.02935} & 953\\
\hline
$\theta$ & MCMC Filter & 0.50118 & 0.44601 & 0.55345 & 92\% & 0.03280 & \textcolor{black}{0.03118} & 947\\
\hline
$\sigma^2$ & MCMC Filter & 1.00734 & 0.93791 & 1.08157 & 88\% & 0.04389 & \textcolor{black}{0.04540} & 956\\
\hline
$d$ & ABC LogP-Q0.005 & 0.17752 & 0.03435 & 0.31391 & 100\% & 0.08522 & \textcolor{black}{0.03719} & 1960\\
\hline
\end{tabular}
\end{table}

\newpage
\begin{table}[H]
\centering
\caption{ARFIMA(0,d,1) Simulation Results for $(d=0.3,\theta=0.2)$}
\centering
\begin{tabular}[t]{l|l|r|r|r|l|r|>{}r|l}
\hline
Par & Model & Mean & LB & UB & Coverage & SD & RMSE & ESS/$\bar{m}$\\
\hline
$d$ & MCMC & 0.29508 & 0.24099 & 0.35265 & 91\% & 0.03406 & \textcolor{black}{0.03323} & 965\\
\hline
$\theta$ & MCMC & 0.20466 & 0.13270 & 0.27355 & 92\% & 0.04295 & \textcolor{black}{\textbf{0.04007}} & 960\\
\hline
$\sigma^2$ & MCMC & 1.00733 & 0.93712 & 1.08133 & 90\% & 0.04400 & \textcolor{black}{0.04531} & 957\\
\hline
$d$ & MCMC Filter & 0.29454 & 0.24057 & 0.35172 & 93\% & 0.03390 & \textcolor{black}{\textbf{0.03289}} & 929\\
\hline
$\theta$ & MCMC Filter & 0.20466 & 0.13287 & 0.27315 & 92\% & 0.04273 & \textcolor{black}{0.04019} & 926\\
\hline
$\sigma^2$ & MCMC Filter & 1.00744 & 0.93773 & 1.08166 & 90\% & 0.04393 & \textcolor{black}{0.04553} & 959\\
\hline
$d$ & ABC LogP-Q0.005 & 0.27424 & 0.13461 & 0.40563 & 100\% & 0.08239 & \textcolor{black}{0.04134} & 1605\\
\hline
\end{tabular}
\end{table}

\newpage
\begin{table}[H]
\centering
\caption{ARFIMA(0,d,1) Simulation Results for $(d=0.3,\theta=0.5)$}
\centering
\begin{tabular}[t]{l|l|r|r|r|l|r|>{}r|l}
\hline
Par & Model & Mean & LB & UB & Coverage & SD & RMSE & ESS/$\bar{m}$\\
\hline
$d$ & MCMC & 0.29616 & 0.24923 & 0.34531 & 92\% & 0.02935 & \textcolor{black}{0.02924} & 965\\
\hline
$\theta$ & MCMC & 0.50261 & 0.44755 & 0.55483 & 92\% & 0.03273 & \textcolor{black}{\textbf{0.03090}} & 962\\
\hline
$\sigma^2$ & MCMC & 1.00720 & 0.93761 & 1.08105 & 89\% & 0.04387 & \textcolor{black}{0.04523} & 961\\
\hline
$d$ & MCMC Filter & 0.29553 & 0.24856 & 0.34470 & 92\% & 0.02932 & \textcolor{black}{\textbf{0.02918}} & 939\\
\hline
$\theta$ & MCMC Filter & 0.50262 & 0.44781 & 0.55454 & 92\% & 0.03260 & \textcolor{black}{0.03096} & 951\\
\hline
$\sigma^2$ & MCMC Filter & 1.00714 & 0.93737 & 1.08127 & 90\% & 0.04394 & \textcolor{black}{0.04521} & 952\\
\hline
$d$ & ABC LogP-Q0.005 & 0.27222 & 0.13842 & 0.39801 & 100\% & 0.07917 & \textcolor{black}{0.04083} & 1863\\
\hline
\end{tabular}
\end{table}

\newpage
\begin{table}[H]
\centering
\caption{ARFIMA(0,d,1) Simulation Results for $(d=0.4,\theta=0.2)$}
\centering
\begin{tabular}[t]{l|l|r|r|r|l|r|>{}r|l}
\hline
Par & Model & Mean & LB & UB & Coverage & SD & RMSE & ESS/$\bar{m}$\\
\hline
$d$ & MCMC & 0.38874 & 0.33705 & 0.44181 & 90\% & 0.03186 & \textcolor{black}{0.03165} & 959\\
\hline
$\theta$ & MCMC & 0.20974 & 0.14050 & 0.27682 & 93\% & 0.04151 & \textcolor{black}{0.03913} & 960\\
\hline
$\sigma^2$ & MCMC & 1.00696 & 0.93745 & 1.08094 & 90\% & 0.04382 & \textcolor{black}{0.04514} & 966\\
\hline
$d$ & MCMC Filter & 0.38527 & 0.33560 & 0.43425 & 90\% & 0.03012 & \textcolor{black}{\textbf{0.03089}} & 945\\
\hline
$\theta$ & MCMC Filter & 0.21293 & 0.14610 & 0.27837 & 93\% & 0.04030 & \textcolor{black}{\textbf{0.03873}} & 943\\
\hline
$\sigma^2$ & MCMC Filter & 1.00696 & 0.93729 & 1.08095 & 89\% & 0.04391 & \textcolor{black}{0.04521} & 957\\
\hline
$d$ & ABC LogP-Q0.005 & 0.36370 & 0.24264 & 0.46940 & 100\% & 0.06926 & \textcolor{black}{0.04671} & 1561\\
\hline
\end{tabular}
\end{table}

\newpage
\begin{table}[H]
\centering
\caption{ARFIMA(0,d,1) Simulation Results for $(d=0.4,\theta=0.5)$}
\centering
\begin{tabular}[t]{l|l|r|r|r|l|r|>{}r|l}
\hline
Par & Model & Mean & LB & UB & Coverage & SD & RMSE & ESS/$\bar{m}$\\
\hline
$d$ & MCMC & 0.39175 & 0.34602 & 0.43858 & 90\% & 0.02816 & \textcolor{black}{0.02815} & 959\\
\hline
$\theta$ & MCMC & 0.50490 & 0.45081 & 0.55630 & 93\% & 0.03221 & \textcolor{black}{0.03052} & 966\\
\hline
$\sigma^2$ & MCMC & 1.00666 & 0.93690 & 1.08048 & 90\% & 0.04381 & \textcolor{black}{0.04500} & 959\\
\hline
$d$ & MCMC Filter & 0.38864 & 0.34433 & 0.43245 & 90\% & 0.02683 & \textcolor{black}{\textbf{0.02731}} & 949\\
\hline
$\theta$ & MCMC Filter & 0.50691 & 0.45407 & 0.55757 & 92\% & 0.03154 & \textcolor{black}{\textbf{0.03024}} & 951\\
\hline
$\sigma^2$ & MCMC Filter & 1.00679 & 0.93711 & 1.08102 & 90\% & 0.04391 & \textcolor{black}{0.04503} & 962\\
\hline
$d$ & ABC LogP-Q0.005 & 0.35989 & 0.23297 & 0.46914 & 100\% & 0.07228 & \textcolor{black}{0.04709} & 2062\\
\hline
\end{tabular}
\end{table}

\newpage
\subsection{Best Performing Method by Bias, RMSE and SD }
\ 
\label{appendix:bestperforming}
\begin{table}[H]
\centering
\caption{ARFIMA$(1,d,1)$ Number of Best Performing Models\label{table:bestperf_arfima1d1}}
\centering
\begin{tabular}[t]{l|r|r|r|r}
\hline
Method & $d$ & $\phi$ & $\theta$ & $\sigma^2$\\
\hline
\multicolumn{5}{l}{\textbf{Bias}}\\
\hline
\hspace{1em}MCMC & 0 & 0 & 1 & 1\\
\hline
\hspace{1em}MCMC Filter & 2 & 1 & 0 & 1\\
\hline
\hspace{1em}ABC LogP-Q0.005 & 0 & 2 & 1 & 0\\
\hline
\hspace{1em}ABC LogP-Q0.01 & 0 & 1 & 0 & 0\\
\hline
\hspace{1em}ABC FP-Q0.005 & 0 & 0 & 1 & 0\\
\hline
\hspace{1em}ABC FP-Q0.01 & 0 & 0 & 3 & 0\\
\hline
\end{tabular}
\end{table}

\begin{table}[H]
\centering
\caption{ARFIMA$(1,d,0)$ Number of Best Performing Model by Bias, RMSE and SD \label{table:bestperf_arfima1d0}}
\centering
\begin{tabular}[t]{l|r|r|r}
\hline
Method & $d$ & $\phi$ & $\sigma^2$\\
\hline
\multicolumn{4}{l}{\textbf{Bias}}\\
\hline
\hspace{1em}MCMC & 0 & 0 & 0\\
\hline
\hspace{1em}MCMC Filter & 1 & 1 & 0\\
\hline
\hspace{1em}ABC LogP-Q0.005 & 0 & 0 & 1\\
\hline
\hspace{1em}ABC LogP-Q0.01 & 0 & 0 & \vphantom{1} 0\\
\hline
\hspace{1em}ABC FP-Q0.005 & 0 & 0 & \vphantom{2} 0\\
\hline
\hspace{1em}ABC FP-Q0.01 & 0 & 0 & \vphantom{2} 0\\
\hline
\end{tabular}
\end{table}

\begin{table}[H]
\centering
\caption{ARFIMA$(0,d,1)$ Number of Best Performing Model by Bias, RMSE and SD \label{table:bestperf_arfima0d1}}
\centering
\begin{tabular}[t]{l|r|r|r}
\hline
Method & $d$ & $\theta$ & $\sigma^2$\\
\hline
\multicolumn{4}{l}{\textbf{Bias}}\\
\hline
\hspace{1em}MCMC & 1 & 0 & 0\\
\hline
\hspace{1em}MCMC Filter & 5 & 3 & 0\\
\hline
\hspace{1em}ABC LogP-Q0.005 & 0 & 1 & 0\\
\hline
\hspace{1em}ABC LogP-Q0.01 & 0 & 0 & \vphantom{1} 0\\
\hline
\hspace{1em}ABC FP-Q0.005 & 0 & 2 & 0\\
\hline
\hspace{1em}ABC FP-Q0.01 & 0 & 2 & 0\\
\hline
\hspace{1em}ABC 20P-Q0.005 & 0 & 1 & \vphantom{2} 0\\
\hline
\hspace{1em}ABC 20P-Q0.01 & 0 & 1 & 0\\
\hline
\hspace{1em}Frequentist & 4 & 0 & 10\\
\hline
\multicolumn{4}{l}{\textbf{RMSE}}\\
\hline
\hspace{1em}MCMC & 1 & 4 & 0\\
\hline
\hspace{1em}MCMC Filter & 7 & 2 & 0\\
\hline
\hspace{1em}ABC LogP-Q0.005 & 0 & 0 & 4\\
\hline
\hspace{1em}ABC LogP-Q0.01 & 0 & 0 & 6\\
\hline
\hspace{1em}ABC FP-Q0.005 & 0 & 1 & 0\\
\hline
\hspace{1em}ABC FP-Q0.01 & 0 & 1 & 0\\
\hline
\hspace{1em}ABC 20P-Q0.005 & 0 & 1 & \vphantom{1} 0\\
\hline
\hspace{1em}ABC 20P-Q0.01 & 0 & 0 & \vphantom{1} 0\\
\hline
\hspace{1em}Frequentist & 2 & 1 & 0\\
\hline
\multicolumn{4}{l}{\textbf{SD}}\\
\hline
\hspace{1em}MCMC & 0 & 1 & 4\\
\hline
\hspace{1em}MCMC Filter & 10 & 9 & 6\\
\hline
\hspace{1em}ABC LogP-Q0.005 & 0 & 0 & \vphantom{1} 0\\
\hline
\hspace{1em}ABC LogP-Q0.01 & 0 & 0 & 0\\
\hline
\hspace{1em}ABC FP-Q0.005 & 0 & 0 & 0\\
\hline
\hspace{1em}ABC FP-Q0.01 & 0 & 0 & 0\\
\hline
\hspace{1em}ABC 20P-Q0.005 & 0 & 0 & 0\\
\hline
\hspace{1em}ABC 20P-Q0.01 & 0 & 0 & 0\\
\hline
\hspace{1em}Frequentist & 0 & 0 & 0\\
\hline
\multicolumn{4}{l}{\textbf{Coverage}}\\
\hline
\hspace{1em}MCMC & 0 & 1 & 3\\
\hline
\hspace{1em}MCMC Filter & 0 & 0 & 3\\
\hline
\hspace{1em}ABC LogP-Q0.005 & 0 & 0 & 0\\
\hline
\hspace{1em}ABC LogP-Q0.01 & 0 & 1 & 0\\
\hline
\hspace{1em}ABC FP-Q0.005 & 3 & 0 & 0\\
\hline
\hspace{1em}ABC FP-Q0.01 & 2 & 2 & 2\\
\hline
\hspace{1em}ABC 20P-Q0.005 & 0 & 1 & 0\\
\hline
\hspace{1em}ABC 20P-Q0.01 & 0 & 2 & 0\\
\hline
\hspace{1em}Frequentist & 9 & 5 & 4\\
\hline
\end{tabular}
\end{table}

\newpage

\twocolumn
\subsection{Forecast Results} \label{appendix:forecast_results}
\ 
\begin{table}[H]
\centering
\caption{ARFIMA$(1,d,1)$ Forecasting Results}
\centering
\begin{tabular}[t]{l|r|r|r}
\hline
Method & RMSE & SD & Coverage\\
\hline
\multicolumn{4}{l}{$d=0.05, \phi=0.2, \theta=0.2$}\\
\hline
\hspace{1em}MCMC & 1.07907 & 1.10374 & 0.89867\\
\hline
\hspace{1em}MCMC Filter & 1.07909 & 1.10244 & 0.90067\\
\hline
\hspace{1em}ABC FP & 1.08008 & 0.98112 & 0.84833\\
\hline
\hspace{1em}ABC 20P & 1.08169 & 1.10231 & 0.89833\\
\hline
\hspace{1em}ABC LogP & 1.07898 & 1.12484 & 0.90567\\
\hline
\hspace{1em}Frequentist & 1.08122 & 1.09656 & 0.89733\\
\hline
\end{tabular}
\end{table}

\begin{table}[H]
\centering
\begin{tabular}[t]{l|r|r|r}
\hline
Method & RMSE & SD & Coverage\\
\hline
\multicolumn{4}{l}{$d=0.2, \phi=0.5, \theta=0.5$}\\
\hline
\hspace{1em}MCMC & 1.75894 & 1.84894 & 0.90100\\
\hline
\hspace{1em}MCMC Filter & 1.76277 & 1.84326 & 0.89867\\
\hline
\hspace{1em}ABC FP & 1.76443 & 1.67204 & 0.86167\\
\hline
\hspace{1em}ABC 20P & 1.76704 & 1.74098 & 0.88267\\
\hline
\hspace{1em}ABC LogP & 1.76557 & 1.82391 & 0.89000\\
\hline
\hspace{1em}Frequentist & 1.76625 & 1.82257 & 0.89767\\
\hline
\end{tabular}
\end{table}

\begin{table}[H]
\centering
\caption{ARFIMA$(1,d,0)$ Forecasting Results}
\centering
\begin{tabular}[t]{l|r|r|r}
\hline
Method & RMSE & SD & Coverage\\
\hline
\multicolumn{4}{l}{$d=0.05, \phi=0.2$}\\
\hline
\hspace{1em}MCMC & 1.01766 & 1.03705 & 0.89867\\
\hline
\hspace{1em}MCMC Filter & 1.01811 & 1.03625 & 0.90133\\
\hline
\hspace{1em}ABC FP & 1.02046 & 0.92868 & 0.85533\\
\hline
\hspace{1em}ABC 20P & 1.01948 & 1.04726 & 0.89900\\
\hline
\hspace{1em}ABC LogP & 1.01779 & 1.05045 & 0.90600\\
\hline
\hspace{1em}Frequentist & 1.01955 & 1.03128 & 0.89633\\
\hline
\end{tabular}
\end{table}

\begin{table}[H]
\centering
\begin{tabular}[t]{l|r|r|r}
\hline
Method & RMSE & SD & Coverage\\
\hline
\multicolumn{4}{l}{$d=0.2, \phi=0.5$}\\
\hline
\hspace{1em}MCMC & 1.22821 & 1.28723 & 0.86167\\
\hline
\hspace{1em}MCMC Filter & 1.18508 & 1.23693 & 0.83000\\
\hline
\hspace{1em}ABC FP & 1.30244 & 1.27824 & 0.88233\\
\hline
\hspace{1em}ABC 20P & 1.30297 & 1.32156 & 0.89467\\
\hline
\hspace{1em}ABC LogP & 1.29986 & 1.34545 & 0.90067\\
\hline
\hspace{1em}Frequentist & 1.30285 & 1.33964 & 0.89900\\
\hline
\end{tabular}
\end{table}

\begin{table}[H]
\centering
\caption{ARFIMA$(0,d,1)$ Forecasting Results}
\centering
\begin{tabular}[t]{l|r|r|r}
\hline
Method & RMSE & SD & Coverage\\
\hline
\multicolumn{4}{l}{$d=0.05, \theta=0.2$}\\
\hline
\hspace{1em}MCMC & 1.01526 & 1.03422 & 0.90000\\
\hline
\hspace{1em}MCMC Filter & 1.01556 & 1.03373 & 0.90100\\
\hline
\hspace{1em}ABC FP & 1.01796 & 0.93087 & 0.86033\\
\hline
\hspace{1em}ABC 20P & 1.01735 & 1.03545 & 0.89800\\
\hline
\hspace{1em}ABC LogP & 1.01492 & 1.04752 & 0.90333\\
\hline
\hspace{1em}Frequentist & 1.01746 & 1.02927 & 0.89833\\
\hline
\end{tabular}
\end{table}

\begin{table}[H]
\centering
\begin{tabular}[t]{l|r|r|r}
\hline
Method & RMSE & SD & Coverage\\
\hline
\multicolumn{4}{l}{$d=0.2, \theta=0.5$}\\
\hline
\hspace{1em}MCMC & 1.20940 & 1.24581 & 0.90000\\
\hline
\hspace{1em}MCMC Filter & 1.20948 & 1.24468 & 0.89733\\
\hline
\hspace{1em}ABC FP & 1.21350 & 1.18151 & 0.87667\\
\hline
\hspace{1em}ABC 20P & 1.21266 & 1.22529 & 0.89367\\
\hline
\hspace{1em}ABC LogP & 1.20913 & 1.23901 & 0.89500\\
\hline
\hspace{1em}Frequentist & 1.21373 & 1.23665 & 0.89133\\
\hline
\end{tabular}
\end{table}

\newpage
\onecolumn

\section{GNP Application} \label{appendix:gnp_application}
\subsection{GNP Model Densities} \label{appendix:gnp_model_densities}
Density plots for ARFIMA models fit on GNP observations. Red lines represent the posterior mean.
\begin{figure}[H]
\begin{center}
\caption{ARFIMA$(1,d,1)$ GNP Simulation Densities} \label{plots:gnp_arfima1d1}
\includegraphics[width=0.9\textwidth]{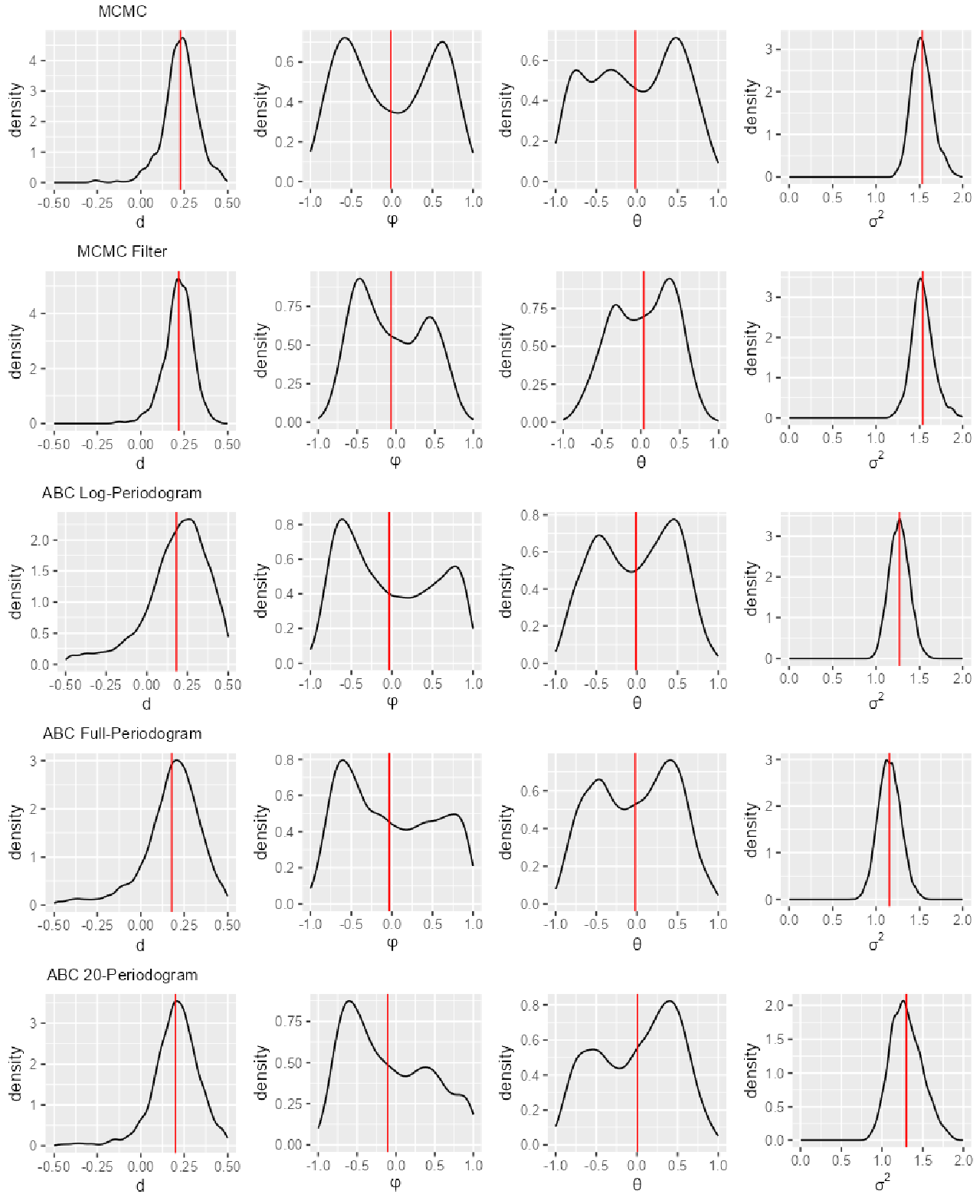}
\end{center}
\end{figure}
\begin{figure}[H]
\begin{center}
\caption{ARFIMA$(1,d,0)$ GNP Simulation Densities} \label{plots:gnp_arfima1d0}
\includegraphics[width=0.7\textwidth]{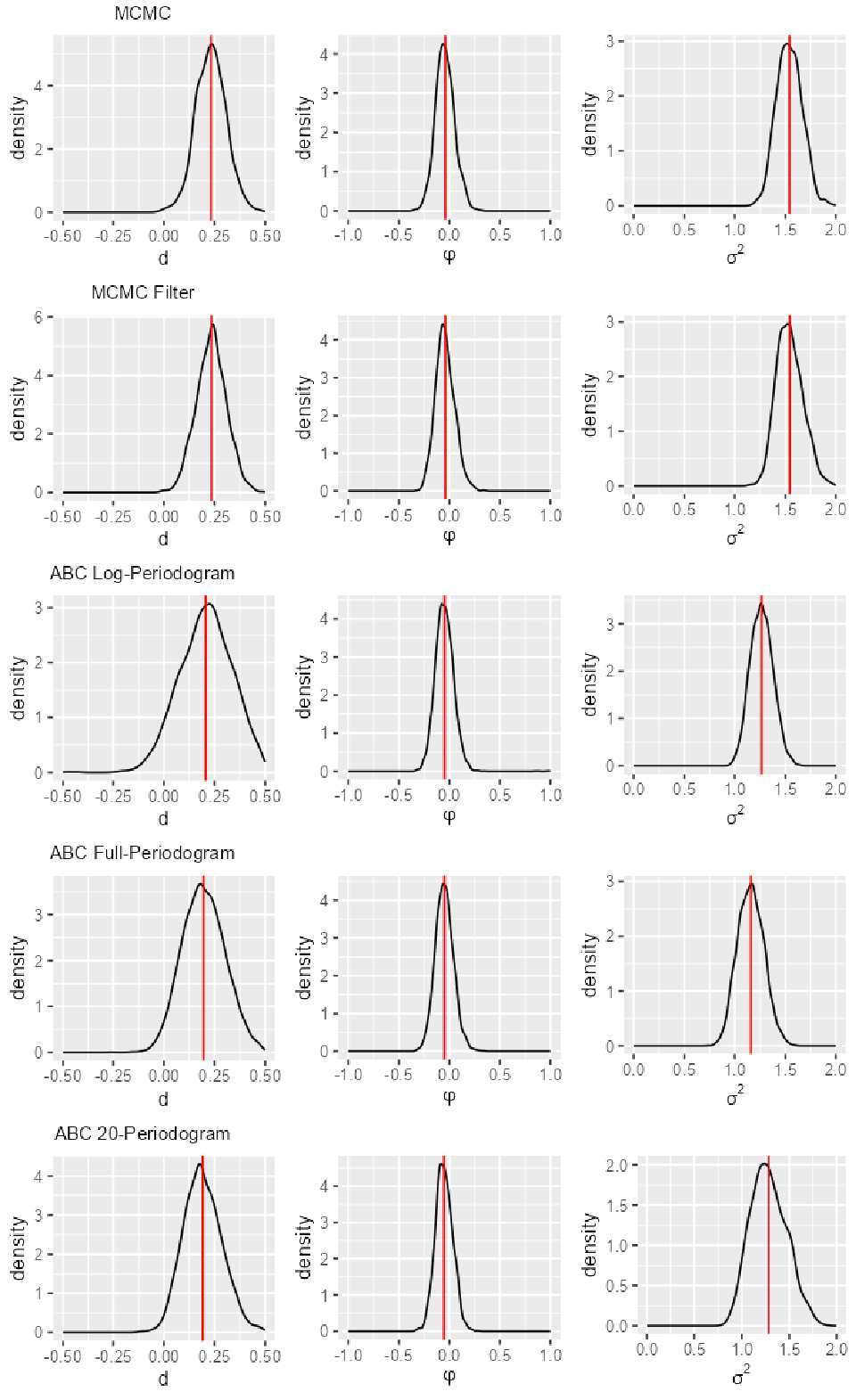}
\end{center}
\end{figure}
\begin{figure}[H]
\begin{center}
\caption{ARFIMA$(0,d,1)$ GNP Simulation Densities} \label{plots:gnp_arfima0d1}
\includegraphics[width=0.7\textwidth]{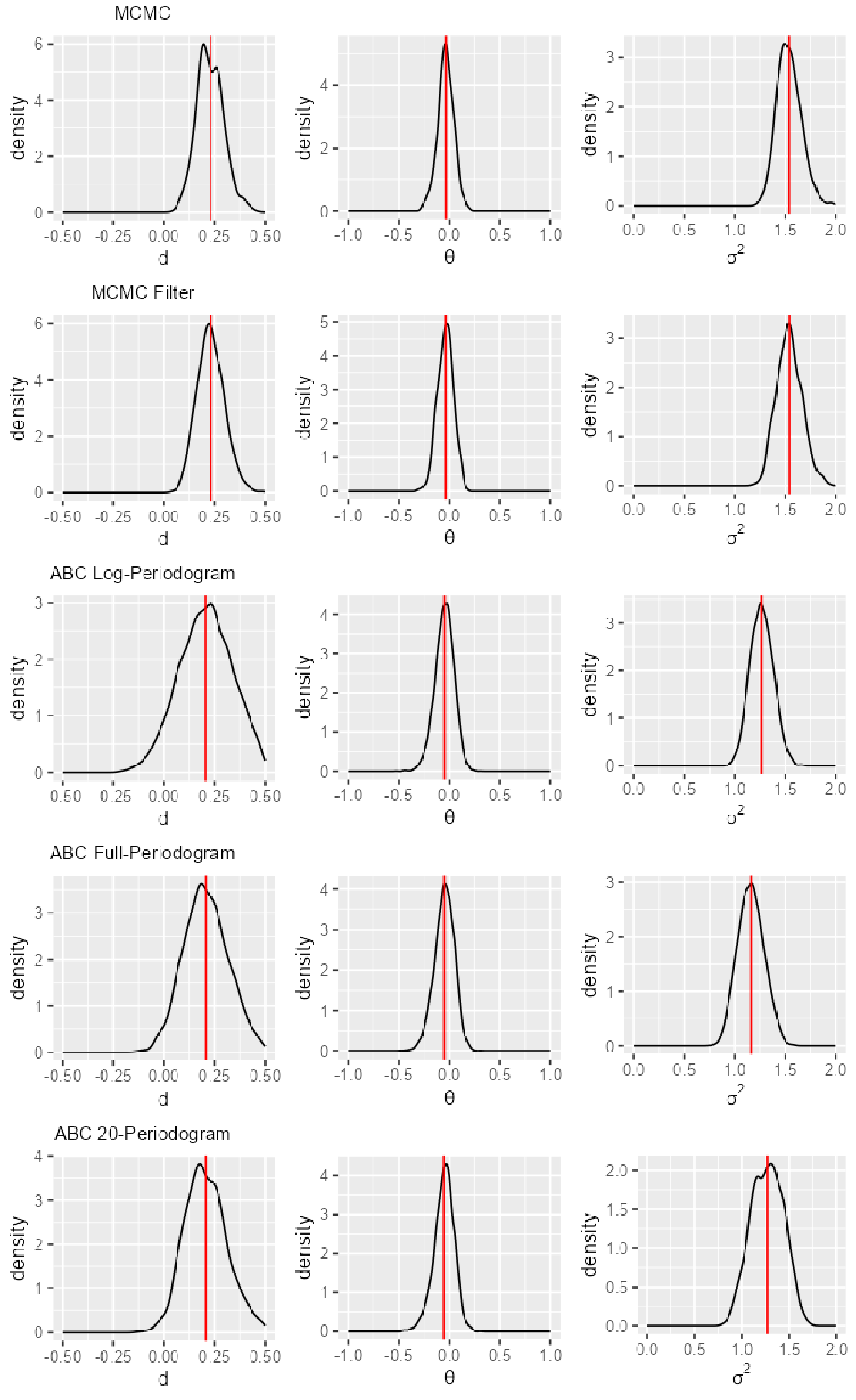}
\end{center}
\end{figure}

\begin{figure}[H]
\begin{center}
\caption{ARFIMA$(0,d,0)$ GNP Simulation Density} \label{plots:gnp_arfima0d0}
\includegraphics[width=0.5\textwidth]{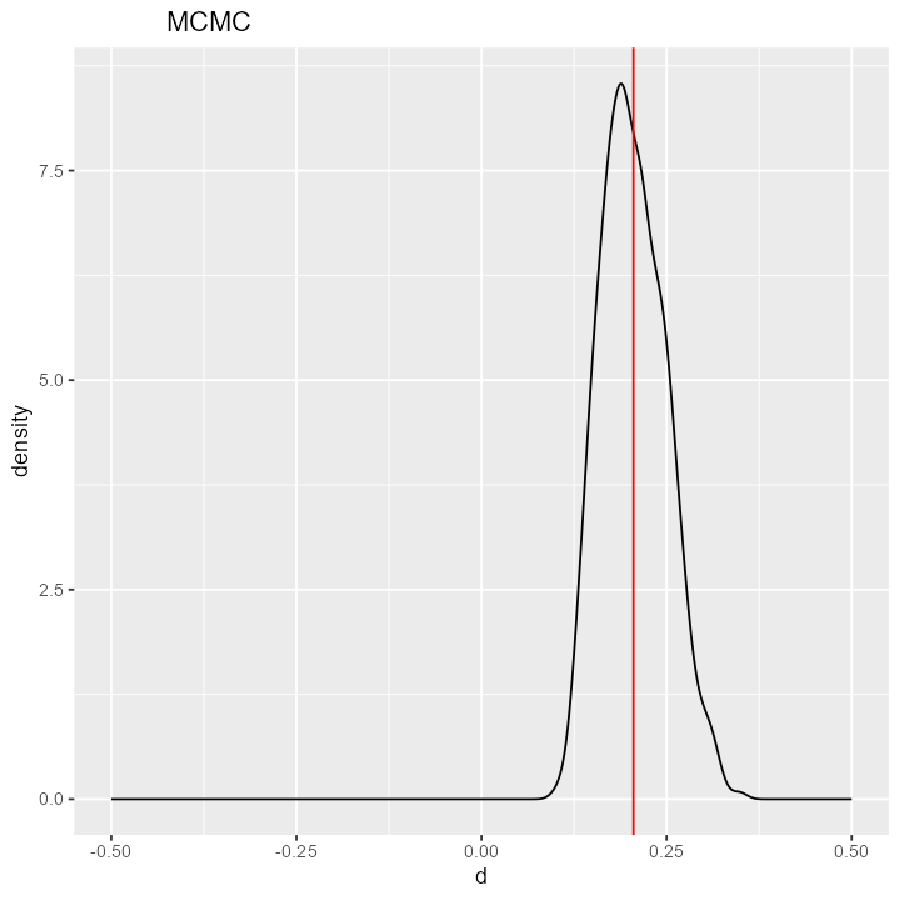}
\end{center}
\end{figure}

\end{appendices}

\end{document}